\documentclass{nature}

\usepackage{graphicx}
\usepackage{setspace}
\usepackage{amsmath}
\usepackage{times}
\usepackage{amssymb}

\usepackage{newtxtext,newtxmath}

\usepackage{times}
\usepackage{sectsty}

\usepackage{lineno}

\title{Earth Shaped by Primordial H$_2$ Atmospheres} 
\author{Edward D. Young,$^{1\ast}$ Anat Shahar,$^{2}$ Hilke E. Schlichting$^1$}

\begin{document} 

\maketitle

\begin{affiliations}
\item{\normalsize{Department of Earth, Planetary, and Space Sciences, University of California, Los Angeles,}}\
\item{\normalsize{Carnegie Institution for Science, Earth and Planets Laboratory, Washington D.C.}}\\
{\normalsize{$^\ast$To whom correspondence should be addressed}\
\normalsize{E-mail:  eyoung@epss.ucla.edu.}}
\end{affiliations}

\begin{abstract}
Earth’s water, intrinsic oxidation state, and metal core density are fundamental chemical features of our planet. Studies of exoplanets provide a useful context for elucidating the source of these chemical traits. Planet formation and evolution models demonstrate that rocky exoplanets commonly formed with hydrogen-rich envelopes that were lost over time \cite{bean2021}. These findings suggest that Earth may also have formed from bodies with H$_2$-rich primary atmospheres.  Here we use a self-consistent thermodynamic model to show that Earth’s water, core density, and overall oxidation state can all be sourced to equilibrium between H$_2$-rich primary atmospheres and underlying magma oceans in its progenitor planetary embryos. Water is produced from dry starting materials resembling enstatite chondrites as oxygen from magma oceans reacts with hydrogen. Hydrogen derived from the atmosphere enters the magma ocean and eventually the metal core at equilibrium, causing metal density deficits matching that of Earth. Oxidation of the silicate rocks from solar-like to Earth-like oxygen fugacities also ensues as Si, along with H and O, alloys with Fe in the cores. Reaction with hydrogen atmospheres and metal-silicate equilibrium thus provides a simple explanation for fundamental features of Earth’s geochemistry that is consistent with rocky planet formation across the galaxy. 
\end{abstract}

 Models for the accretion of the Earth date back decades \cite{Wetherill1978}. Modern versions often focus on matching Earth's chemical features by ascribing different chemical characteristics to possible planetary embryos and planetesimals in N-body simulations of the accretion process. In order to match Earth's chemistry, gradients in parameters such as water availability and oxygen fugacity across the protoplanetary disk are invoked along with prescriptions for partial equilibration during discrete collision events \cite{Rubie2015}. These models by necessity involve a large number of adjustable parameters, thus permitting reasonable fits.  The models are logically sound, but their veracity is less clear.  We present an alternative approach that accounts for recent insights gained from rocky exoplanet formation and the important role of primary atmospheres, and use this to re-examine Earth's formation. 
 
 Two properties of the inner Solar System would seem to have worked against the formation of a water-rich, oxidized (relative to a solar gas) planet like the Earth: 1) the material in the vicinity of 1 AU from our star was likely dry \cite{Albarede2009}, and 2) this material was relatively reduced (i.e, not oxidized) \cite{Cartier2019}. These conclusions are based on the emerging view that the best samplings of inner Solar System rocks, apart from Earth itself, are enstatite chondrites (E chondrites) and their igneous equivalents (aubrites) \cite{Dauphas2017}.  E chondrites closely resemble the Earth in at least 15 isotopic systems \cite{Sikdar2020} and, unlike other meteorite groups, have sufficient metal to explain the mass fraction of Earth's core \cite{Javoy1995}. What is more, the inner Solar System is bound, spatially, by E-chondrite like material.  The inner edge is demarcated by Mercury, which is compositionally similar to E chondrites and aubrites \cite{Nittler2011}. The outer edge is marked by the inner-most  asteroid belt, where E-type asteroids are most abundant; E-type and some M-type asteroids are likely sources for E-clan meteorites \cite{Shepard2015,Zellner1977}.  This further suggests that the inner Solar System was mainly, although not entirely \cite{Piani2020}, dry and reducing much like E-clan meteorites. Enstatite chondrites themselves may represent olvine-poor vestiges of Earth's building blocks \cite{Dauphas2017}.      

Recent exoplanet observations have revealed that the most common planets in our Galaxy discovered to date are larger than Earth and smaller than Neptune \cite{Petigura2013b}, and comprise two distinct groups.  One group is composed of rocky planets with hydrogen-rich envelopes that make up a few percent of their total mass, the so-called sub-Neptunes. The other includes rocky planets absent H$_2$-rich atmospheres, the so-called super-Earths \cite{weiss2014a}. These two populations are divided by a ``radius valley" in planet-size near $1.5$ to $2.0$ $R_{\oplus}$ \cite{fulton2017a}. Planet formation and evolution models to explain the radius valley have demonstrated that the rocky super-Earths we observe today likely formed with hydrogen-rich envelopes that were lost over time \cite{Berger2020} (in about $10^8$ to $10^9$ years) by either photo-evaporation and/or core-powered mass loss, suggesting that the super-Earths and sub-Neptunes originally formed as one population with H$_2$ atmospheres \cite{owen2013a, gupta2019a}. Through the studies of these exoplanets, we have gained a greater appreciation for the ubiquity of H$_2$-rich atmospheres accreted from protoplanetary disks during the first several million years of planetary growth \cite{ginzburg2016a}. Our calculations illustrate how this process may have shaped Earth's chemistry and that Earth's formation can be successfully placed into the context of rocky exoplanet formation within our Galaxy.

While decay of the short-lived radionuclide $^{26}$Al dominated the thermal structure of small planetesimals formed early in the Solar System, the thermal histories of more massive bodies ($> \sim 0.05 M_\oplus$ ) were controlled by the conversion of gravitational potential energy into heat during accretion. As a result, planetary embryos in the early Solar System were initially molten. Upon cooling to below the threshold surface temperature for thermal escape of gas molecules, the embryos are expected to have accreted gas from the surrounding protoplanetary disk, thus acquiring hydrogen-rich primary atmospheres.  Once an optically thick envelope has been accreted, subsequent cooling of the embryo-atmosphere system occurs through the top of the atmosphere at the radiative-convective boundary ($R_{\rm RCB}$) \cite{ginzburg2016a}.  The cooling is slowed relative to radiation from the surface of the magma ocean by the factor $(T_{\rm e}^4/\tau)/T_{s}^4$ where $T_{\rm e}$ is the equilibrium temperature, $\tau$ is the Rosseland mean optical depth at the $R_{\rm RCB}$, and $T_{s}$ is the surface temperature of the magma ocean \cite{ginzburg2016a}. For the relevant optical depths at the radiative-convective boundary of atmospheres considered here (e.g., $\tau \sim 100$) and equilibrium temperatures similar to Earth's, cooling is slower by a factor of $\sim 10^6$ relative to the no-atmosphere case, resulting in magma oceans below the atmospheres that persist for tens of millions of years.  

If material and chemical exchange is efficient, chemical equilibrium between the magma oceans and the atmospheres will occur \cite{Schlichting_Young_2022}.  Convective velocities in the magma ocean are the limiting factor for the rate of exchange.  Scaling laws based on experiments indicate that convection in magma oceans the size of the embryos leads to linear velocities of order $1$ m/s for surface temperatures above the solidus \cite{Solomatov2009,Young2019}.  The effect of the optically-thick atmosphere is to decrease the radiative flux and thus slow the convective velocity by the factor $ ((T_{\rm e}^4/\tau) /T_{s}^4)^{1/3} \sim 100$. The resulting convective turnover times of order $10^4$ yr are still more than sufficient to allow equilibrium between the magma oceans and the overlying atmospheres over the  $\sim 10^7$ yr cooling timescales of the magma ocean-atmosphere system.    

Ingress of protoplanetary disk gas (solar gas) into Earth's building blocks is evidenced by noble gas isotope ratios from mantle rocks \cite{Mukhopadhyay2012,Mukhopadhyay2019,Williams2019,Lammer2020,Sharp2022}. Conversely, it appears that Mars has no vestiges of solar gas in its interior \cite{Kurokawa2021,Peron2022}. Because gas is soluble in molten rock but not solid rock, the lack of solar gas in the interior of Mars is to be expected. Isotopic constraints as well as the similarity of its mass to the so-called ``isolation" mass for a minimum-mass solar nebula (MMSN), suggest that Mars is a planetary embryo that grew within about 2 to 4 Myr and likely had a magma ocean \cite{Dauphas2011}.  However, Mars was insufficiently massive to retain an atmosphere at surface temperatures consistent with molten rock, thus precluding exchange between the atmosphere and the interior (see Methods). Comparing the gravitational binding energy to the thermal energy of hydrogen gas molecules implies that a body the mass of Mars, $0.1 M_\oplus$, can  only accrete a primary hydrogen-dominated atmosphere from the disk after its surface cooled to $\le 700$ K, well below the solidus for rock (see Extended Data Figure 1) \cite{ginzburg2016a}. Therefore, one would expect that a magma ocean would not have been exposed to the primary atmosphere in the case of Mars, consistent with isotopic constraints.

Not all planetary embryos that grew in the presence of the protoplanetary disk need have been the mass of Mars, however. Surface densities greater than that of the MMSN, radial drift of disk material \cite{schlichting2014a}, radial migration of embryos themselves, and/or rapid growth facilitated by pebble accretion \cite{Johansen2021}, would have resulted in growth of embryos to several tenths of Earth masses.  These bodies would have retained substantial primary atmospheres of hydrogen with surface temperatures $> 2000$ K, well above the solidus \cite{ginzburg2016a}. The threshold mass for accretion of an H$_2$-rich atmosphere with surface temperatures above the silicate solidus of $\sim 1500$ K is about $0.2 M_\oplus$. As discussed above, once an embryo has accreted an optically thick hydrogen envelope, these temperatures can persist for as long as an optically thick atmosphere is present.

We explored the consequences of chemical exchange between hydrogen atmospheres and magma oceans for the chemical evolution of planetary embryos that formed the Earth. The model is composed of 25 phase components, including  MgSiO$_3$, MgO, SiO$_2$, FeSiO$_3$, FeO, Na$_2$SiO$_3$, Na$_2$O, H$_2$, H$_2$O, CO and CO$_2$  species in silicate melts, Fe, Si, O, and H in metal melts, and H$_2$, CO, CO$_2$, CH$_4$, O$_2$, H$_2$O,  Fe, Mg, Na, and SiO in the atmospheres.  This system spans a reaction space of 18 linearly independent reactions that account for speciation in the magma ocean, exchange between the atmosphere and the magma ocean, and exchange between silicate and metal in the magma ocean.  For each of the 18 independent reactions we solved the conditions for thermodynamic equilibrium together with the mass balance constraints for the atmosphere, silicate melt, and metal melt, as well as the pressure at the base of the atmosphere as prescribed by the mean molecular weight of the atmosphere and gravitational acceleration (see Methods).  Solutions were obtained by a combination of simulated annealing and Markov chain Monte Carlo (MCMC) sampling (see Methods), as described previously by Schlichting and Young \cite{Schlichting_Young_2022}.  This chemical system is meant to be simple enough to interpret but sufficiently complex to account for the salient features of atmosphere-magma ocean exchange. 

While we assumed ideal mixing for the atmosphere and silicate melt (see Methods), precise  concentrations of light elements in the metal are a focus of our study, so in this case we used the non-ideal mixing model for the  alloying species O, Si, and Fe given by Badro et al. \cite{Badro2015}. The thermodynamic mixing behavior of H in iron metal is not as well characterized. However, ideal mixing as a first approximation is suggested by the small size of H atoms forming interstitial alloys with Fe. Recent results \cite{Li2020_1038} showing that H bonds mainly with Fe in metal with no preference for bonding with O lend support for this approximation. The effects of non-ideal mixing of H in the metal alloy are described in the Methods section. 

We used a fiducial embryo mass of $0.5 M_\oplus$ for illustration, although our results are not critically dependent on the precise masses of the embryos as long as they exceed $\sim 0.2 M_\oplus$. The initial composition of the body is composed of an Fe metal fraction of $34.4 \%$, resulting in a final metal alloy mass fraction of $32.5 \%$ in our models, consistent with Earth's core fraction.  The initial silicate is consistent with estimates for the bulk silicate Earth (BSE) projected into the model composition space composed of Mg, Si, Fe, Na, and O (i.e., excluding Ca and Al).  We assumed the initial concentration of total oxidized iron, where all iron in the silicate is cast as FeO for reporting purposes, to be  $0.07$ wt \% to simulate the reduced nature of the inner solar system as evidenced by E chondrites and Mercury ($\Delta{\rm IW} \sim -5$ compared with $\Delta{\rm IW} \sim -2$ for Earth, see below). 

We performed calculations over a range of metal-silicate equilibration temperatures and corresponding potential temperatures at the top of the magma ocean (Figure 1).  A metal-silicate equilibration temperature of 3000 K was adopted as our fiducial model as this is the temperature indicated by element partitioning between Earth's silicate and metal at pressures of about 40 GPa for single-stage and multi-stage equilibration models involving progressively increasing oxygen fugacities \cite{Wood2008}. This allows for either inheritance of metal-silicate elemental partitioning from Earth's progenitor planetary embryos, or, subsequent reequilibration following, for example, the giant impact that formed the Moon, as suggested by the Hf-W isotopic system \cite{Wood2008}. Based on a Vinet equation of state (EOS) for the molten metal core  \cite{seager2007a,Anderson2001, Kuwayama2020, Schlichting_Young_2022} and a third-order Birch-Murnaghan EOS for the silicate \cite{seager2007a}, the pressure at the core-mantle boundary of a fully differentiated $0.5 M\oplus$ body with Earth-like fractions of metal is $\sim 60$ GPa (see Methods), suggesting the possibility of equilibration at depths shallower than the core-mantle boundary in these models (as in the case of Earth itself).  Embryo masses of $0.3 M\oplus$ result in a core-mantle boundary pressure of 40 GPa.  For our fiducial case, the potential temperature of the magma ocean of 2350 K is used for the surface of the magma ocean, reflecting the thermal insulating effect of the dense, optically thick H$_2$ atmospheres. The initial pressure of the primary atmosphere in our favored solution is $0.13$ GPa (1338 bar), corresponding to a total H-mass fraction of about 0.2\%, which is within the range of predicted hydrogen mass fractions from atmospheric accretion models \cite{ginzburg2016a}. The potential temperature varies by  approximately 100 degrees depending on  whether the 3000 K equilibration occurs somewhat above the final core-mantle boundary at 40 GPa or at the final core-mantle boundary at 60 GPa.  Our results are not sensitive to this uncertainty in pressure.

Results show that the overall effect of reaction between magma oceans and H$_2$-rich primary atmospheres is transfer of large masses of hydrogen to the metal phase, oxidation of the atmosphere, and production of significant quantities of water (Figure 1). The production of water by reactions between H$_2$ atmospheres and the underlying planet has been pointed out previously \cite{Ikoma2006,Kite2021}.  Here we find that the conditions for production of water by the coupled equilibria are accompanied by incorporation of hydrogen into the metal phase. Light elements in the metal cores 
 and total water are positively correlated, illustrating that water is a by-product of the redox reactions that drive hydrogen into the metal. Water is also a product of oxidation of the atmosphere by evaporation of oxides comprising the silicate melt \cite{Schlichting_Young_2022}. Water produced in the interiors of the embryos and in the atmosphere partitions according to the solubility of water in the magma ocean.  Partitioning between melt and atmosphere is sensitive to the activity-composition relationship for water in the melt.  Dissolution of water as OH shifts more water to the silicate melt than shown in Figure 1B.

Both silicon and oxygen are also incorporated into metal as the embryos equilibrate in these calculations, contributing to the density deficits in the metal relative to pure Fe.  Earth is known to have a density deficit relative to pure Fe (+Ni) of $\sim 10\, \%$ \cite{Birch1964}.  The Birch-Murnaghan equation of state for Fe-H alloy determined by Umemeto and Hirose \cite{Umemoto2020} shows that the density deficit caused by H at 3000 K and 40 GPa increases by up to $25\, \%$ when compressed to 136 GPa at temperatures $> 4000$ K, the conditions for Earth's outermost core. Therefore, a deficit in metal density of $8\, \%$ relative to pure iron in the embryos  as a result of reactions with H$_2$-rich primary atmospheres could  produce a $10\, \%$ deficit in Earth's core upon compression. An $8\, \%$ density deficit is obtained in our fiducial model for a total H concentration of $0.2\, \%$ by mass (Figure 1A).  In this case, the calculated metal cores are composed of $94.9\, \%$ by mass Fe, $3.8\, \%$ Si, $0.8\, \%$ O, and $0.5\, \%$ H. Because a density deficit relative to pure iron of $8.7\, \%$ results for each weight percent of H in Fe alloys, while the values for O and Si are $1.2$ and $0.8\, \%$, respectively \cite{Li2019}, $3/4$ of the density deficit in the cores is attributable to H. For comparison, a $10\, \%$ density deficit in embryo metal is obtained from a body composed of $0.3\%$ by mass H (Figure 1A). At this total hydrogen content, the calculated metal cores are composed of $94.6\, \%$ by mass Fe, $3.8\, \%$ Si, $0.9\, \%$ O, and $0.7\, \%$ H.   

 The H mass fraction of $\sim 0.2\, \%$ that corresponds to an embryo metal density deficit of $8\, \%$ also produces a mass fraction of water corresponding to approximately one terrestrial ocean  (Figure 1B). For a total H mass fraction of $0.3\, \%$, corresponding to a $10\, \%$ density deficit in the embryo cores, the fractional amount of water produced is about 2 to 3 ocean equivalents. These results are similar to that of an earlier study on production of water by oxidation of atmospheric H$_2$ \cite{Ikoma2006}. Even if the rocks in the inner Solar System were entirely dry, reactions between H$_2$ atmospheres and magma oceans would generate copious amounts of water.  Other sources of water are possible, but not required.
 
Earth's intrinsic oxidation state as recorded by the amount of FeO in the BSE derived from our calculations is consistent with the observed value. This is a measure of self consistency between the oxidation state of the planet and the composition of the metal core afforded by this model. The oxidation state of silicates in planetary bodies is usually reported in terms of oxygen fugacity, or the non-ideal partial pressure of oxygen that would be defined by equilibrium with the silicate if O$_2$ were exchanged between a vapor phase and the melt. By convention, oxygen fugacities are reported relative to the reaction between pure iron and pure FeO (w\"{u}stite) according to the reaction Fe  + $1/2$ O$_2$ = FeO as $\Delta {\rm IW} = 2\log(x_{\rm FeO}/x_{\rm Fe})$ where $x_{\rm FeO}$ is the mole fraction of FeO (as total iron) in the silicate and  $x_{\rm Fe}$ is the mole fraction of Fe in metal, assuming ideal mixing (again, by convention).  On this scale, E chondrites and aubrites have $\Delta {\rm IW}$ values of $\sim -4.5$ to $-6.5$, similar to a hydrogen-rich solar gas, and Earth has a value of $\sim -2.2$ \cite{doyle2019a} (Figure 2).  We used an E chondrite value of $-5.8$ as our initial condition.  Equilibration of the system yields an Earth-like $\Delta {\rm IW}$ of $\sim -2.1$ as a result of oxidation of Fe displaced from the core by Si, a process that can be written as  Si$^{4+}$ + 2Fe$^0$ = Si$^0$ + 2Fe$^{2+}$ that has been suggested previously as a mechanism for oxidizing the mantle \cite{Javoy2010,Rubie2015}. While this process can occur in the absence of hydrogen, our calculations show that the Earth's intrinsic oxidation state is consistent with its core density that is in turn attributable to equilibration with an H$_2$ atmosphere. This need not have been the case. 

The two adjustable parameters in our calculations, core-mantle temperature of equilibration and the initial mass fraction of H$_2$, are degenerate with respect to the density of the core.  However, in order to fit the core density with higher temperatures and substantially less H, the density deficit must come from more and more Si and O in the core (arrow in Figure 1A). More Si ( $> 6$ wt $\%$, see Extended Data Figure 2) in the core results in a higher intrinsic oxygen fugacity by several tenths of log units (arrow in Figure 1C), reaching values greater than estimates for the Earth.  In addition, little to no water is formed in the absence of a significant primary H$_2$ atmosphere (arrows in Figure 1B), requiring in this case that Earth's water must be entirely exogenous.  

Our prediction that the primary light element in Earth's core that is responsible for the density deficit relative to pure Fe and Ni is H has implications for the  seismic velocities ($V_{\rm P}$) in the outer core. The outer core has a higher $V_{\rm P}$ relative to pure Fe by several percent \cite{Dziewonski1981}.  Both H and Si increase $V_{\rm P}$ in molten iron alloys, with the small concentrations of oxygen in our models having negligible effects.  The largest effect is expected for H \cite{Sanloup2004,Umemoto2015}.  Umemeoto and Hirose \cite{Umemoto2015} showed that for Fe-H alloys similar in composition to those in our models, $V_{\rm p}$ increases by about $1.8\, \%$ relative to pure Fe-Ni alloys, and yields values for $V_{\rm P}$ consistent with the  value for the the outermost core \cite{Kennett1995}. The outer core densities and seismic velocities implied by our model core composition are within about 2 \% of measured values based on the published equations of state \cite{Umemoto2020, Badro2015}, which we consider a reasonable fit given differences between models based on different equations of state. 

Figure 3 summarizes our proposed model for the evolution of Earth's progenitor embryos with masses on the order of a few tenths of $M_\oplus$.  Our fiducial  calculation should be regarded as a surrogate for the average of what may be several embryos that eventually combined to form Earth.  Equilibration between metal and silicate in planetary embryos provides a natural explanation for the relatively low pressures of $\sim 40$ GPa attending element partitioning for Earth compared with Earth's core-mantle boundary pressure of 136 GPa as this lower pressure corresponds to those near the core-mantle boundaries of embryos. The precise masses of Earth's progenitor embryos are not crucial to our conclusions. For example, our modeling results are virtually identical using the smaller embryo mass of $0.3 M_{\oplus}$ where 3000 K and 40 GPa represents the core-mantle boundary.  In this case, the potential temperature is 2535 K, but nonetheless the core composition, water fractions, oxygen fugacity, and other model results are unchanged (see Supplementary Data Tables).

If this is the explanation for the low pressures of equilibration, core-mantle partitioning was inherited from the embryos from which Earth was built, and many of these chemical signatures survived subsequent collisions among embryos, including the Moon-forming giant impact.  The alternative is that melting during giant impacts reset core-mantle equilibrium at similar conditions \cite{Wood2008}.  Collisions between embryos after the protoplanetary gas disk has dissipated would efficiently remove any residual hydrogen dominated atmospheres.  However, the water-rich atmospheres left behind by equilibrium between the H$_2$ primary atmospheres and magma oceans will be largely retained \cite{biersteker2021a}. As a result, most of the water produced in the embryo stage will remain once the disk dissipates \cite{biersteker2019a}. In order to test the effects of colliding two embryos with residual water atmospheres to produce the Earth, we applied our model to an initial condition of a $1 M_\oplus$ body with mantle, core, and water-rich atmosphere compositions corresponding to our final embryo equilibrium state.  Results verify that the chemical signature of hydrogen on embryos survives subsequent impact-generated re-equilibration at similar temperatures.

In the context of our model for Earth formation involving a primary atmosphere of hydrogen,  both Mars and Mercury are planetary embryos that never reached sufficient mass to allow primary atmospheres of H$_2$ at magma ocean surface temperatures. The identities of the light elements in the core of Mars are not known with certainty, but are generally believed to be dominated by sulfur \cite{Stahler2021}.  The effect of S on the density deficit of metal alloys is similar to that of Si, and tens of per cent by mass S are required to account for the density of the core. Also in the context of this model, the relatively high intrinsic oxidation state of Mars (high FeO in silicate, but see also \cite{Stahler2021}) may be attributable to a large fraction of its mass being derived from beyond 2 AU where the more oxidized and water-rich carbonaceous chondrites could serve as sources \cite{Brasser2013}. In the case of Mercury, if one assumes that the large metal fraction of Mercury is due to stripping of its mantle \cite{Benz2007}, and that the original mass fraction of its metal core was similar to that of Earth (0.32), then the mass of the original body was between $0.09 M_\oplus$ and $0.13 M_\oplus$ for a range of estimated present-day core fractions of 55 to 75\% by mass \cite{Riner2008, Margot2018}.  Like Mars, this is not sufficiently massive to retain an H$_2$ atmosphere at super-solidus temperatures. The light element in the Mercurian core is thought to be mainly Si, with lesser concentrations of S and no clear evidence for H.  Therefore, the chemical distinction between Earth and these stranded embryos may in part be that Earth was built from embryos that reached sufficient mass to retain H$_2$-rich atmospheres at super-solidus surface temperatures.

Our model does not include sulfur and we do not exclude the possibility that the details of the model might change if sulfur were included.  Similarly, we do not include carbon or nitrogen as potential interstitial alloying elements in the core. Omission of Ni is not expected to have a significant effect. While details of our conclusions may change, the essential point that reactions with hydrogen-rich atmospheres are effective in producing both water and density deficits in metal cores remains robust, as does the result that the silicate mantles of  progenitor embryos resembling E chondrites and Mercury in having low total FeO, should have ``self oxidized" to Earth-like oxidation states given sufficient time to approach equilibrium \cite{Javoy2010}.  We provide a discussion of the isotopic consequences of our results in the Methods section.   

This study makes use of what we have learned from rocky exoplanets about the potential importance of primary H$_2$-rich atmospheres to terrestrial planet formation in general.  The result is a unified, self-consistent explanation for a number of important features of Earth related by a single overarching process of oxidation and reduction (redox) chemistry triggered by reactions between H$_2$ and magma oceans. While it is possible that Earth formed in a very different way from the majority of rocky exoplanets, our work shows that this need not be the case and that we can place the formation of Earth into the context of rocky planet formation across the Galaxy.

\begin{singlespace}

\noindent 
 \begin{figure}[!h]
\centering
\includegraphics[width=5.8 in]{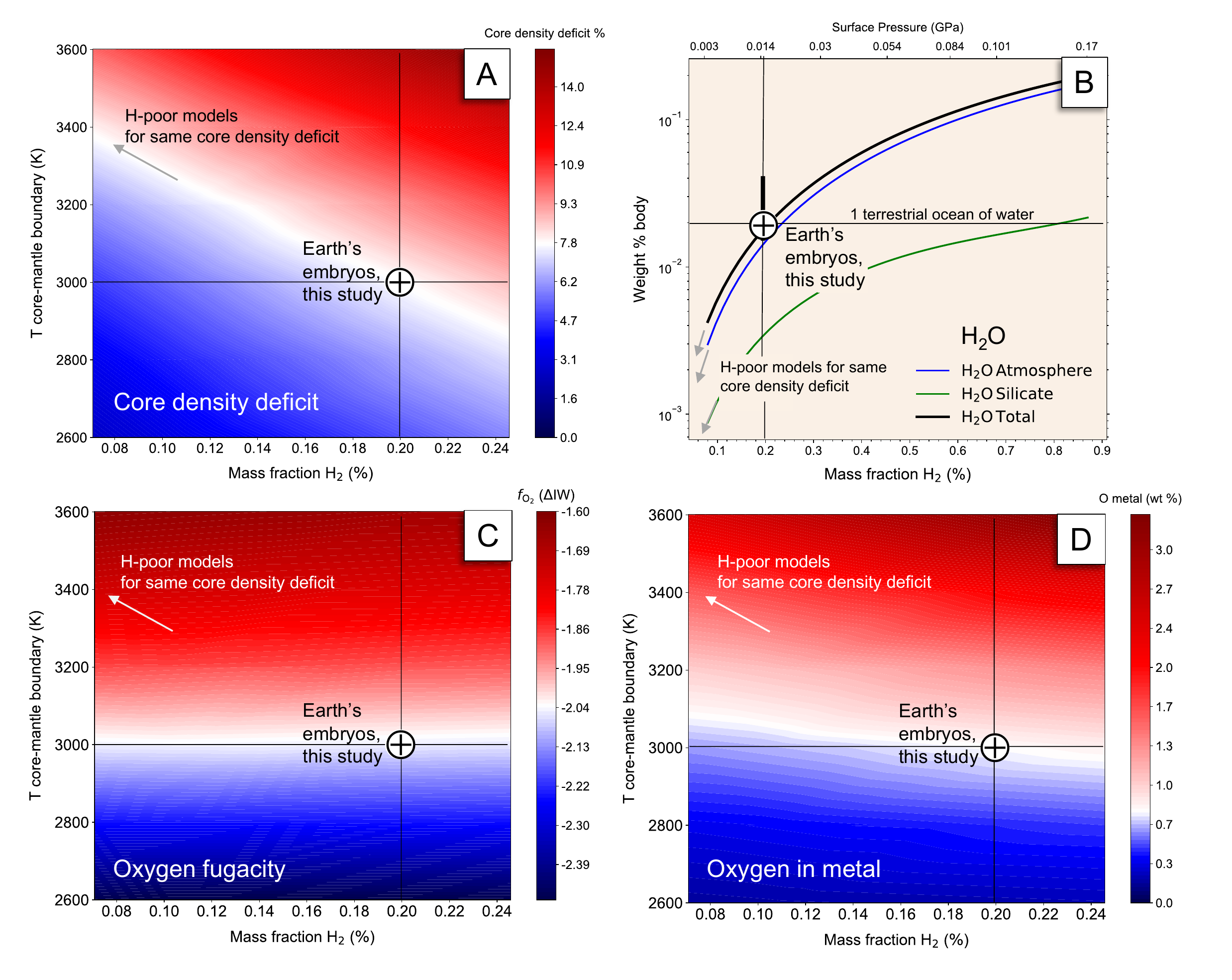}
\caption{Summary of thermodynamic calculations showing that  Earth's core density deficit (A), mass fraction of water (B), and oxygen fugacity (C) can be explained in a self-consistent way with a global equilibrium model that includes an H$_2$-rich primary atmosphere, a silicate magma ocean and a metal core. Results are shown for $0.5 M_\oplus$ embryos as a function of metal-silicate equilibration temperature ($T_{\rm core-mantle}$) and the initial mass fraction of primary H$_2$-rich atmospheres. In each panel parameters for embryos that would fit the Earth at $T_{\rm core-mantle} = 3000$ K (see text) are shown with the $\oplus$ symbol. The target values for Earth are shown in the white regions in the contour plots.  Panel A shows density deficits in metal cores with the target value of $\sim 8\%$ by weight (see text).  Panel B shows the  H$_2$O mass fraction for the embryos where $T_{\rm core-mantle} = 3000$ K showing that the model satisfying Earth's core density deficit (A) results in one Earth's oceans worth of water ($\oplus$).  Here the pressures at the base of the atmosphere after equilibration are shown on the upper abscissa, with the best-fit for Earth's antecedent embryos corresponding to pressures at the surface of the magma ocean of $\sim 0.014$ GPa (140 bar). Panel C shows the intrinsic oxygen fugacities relative to the iron-w\"{u}stite reference, illustrating that the same model that matches Earth's core density yields Earth's intrinsic oxygen fugacity of $\sim \Delta{\rm IW} -2.1$. Panel D shows the weight percents of O in metal as predicted by this model. Arrow in panel A points to the direction of models that satisfy Earth's core density deficit but where H is scarce or absent. This same trajectory is illustrated in panels B, C, and D. }
\end{figure}

\begin{figure}[!h]
\centering
\includegraphics[width=3.8 in]{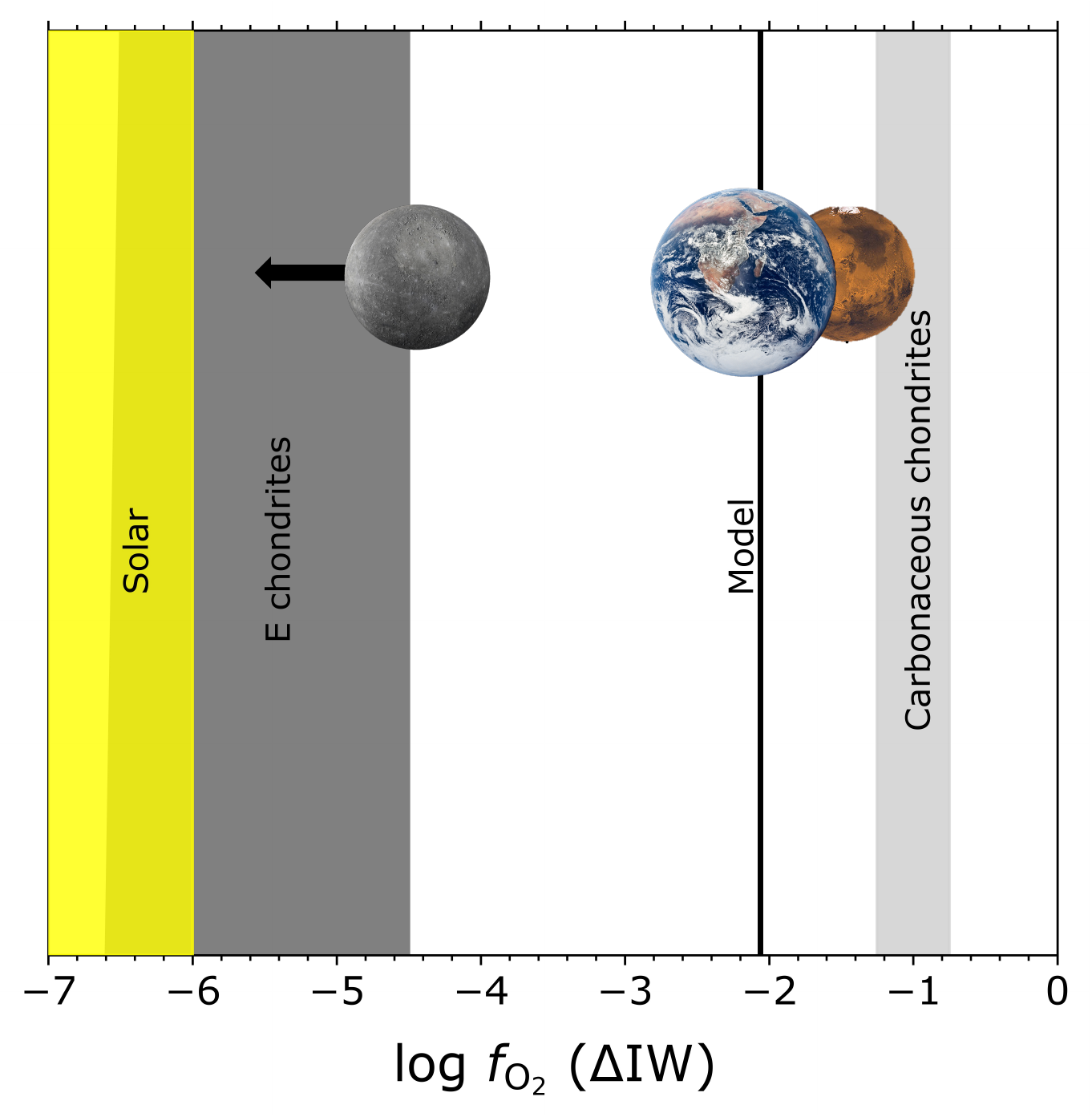}
\caption{The model for equilibration between magma oceans and H$_2$-rich atmospheres (vertical black line) results in an increase in oxygen fugacity from values similar to Mercury and E chondrites, representing the inner solar system, to the value for Earth. This figures compares the model oxygen fugacity for embryos matching Earth with values from the Solar System, including those for a solar gas, E chondrites, carbonaceous chondrites, Mercury, Earth, and Mars. All values are intrinsic oxygen fugacities based on bulk-silicate FeO concentrations \cite{doyle2019a} and are reported as log$f_{\rm O_2}$ relative to the iron-w\"{u}stite buffer. The value for Mercury is actually a range from about $-4$ to $-6$, depending on the study. } 
\end{figure}

\begin{figure}[!h]
\centering
\includegraphics[width=5.8 in]{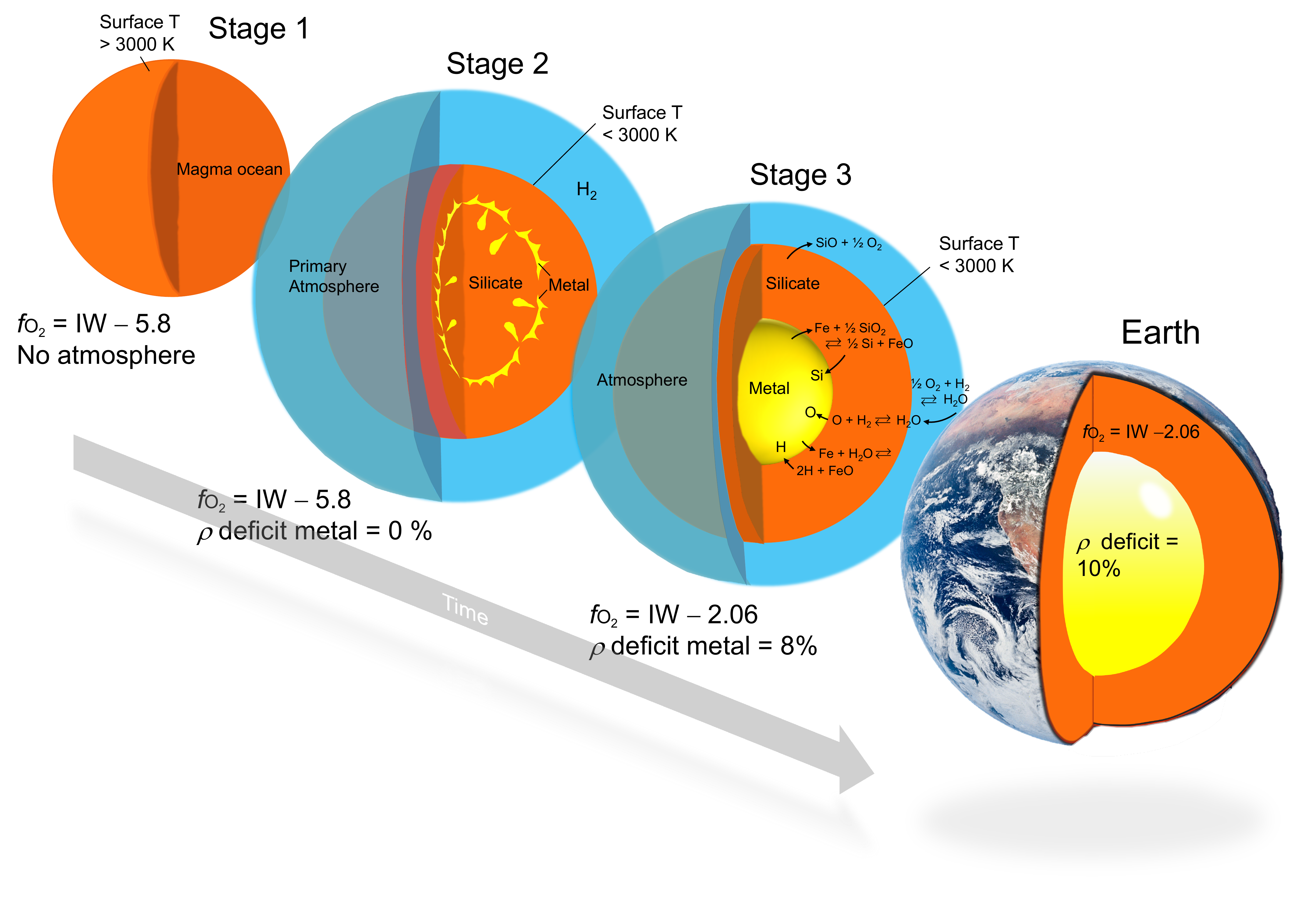}
\caption{The sequence of events leading to formation of water, light elements in metal, and increases in oxygen fugacity for Earth's progenitor embryos in this work. Stage 1 is the embryo where surface temperatures are too high to retain primary atmospheres of hydrogen.  Stage 2 is the initial condition for our calculations in which the molten embryos accrete and retain primary atmospheres of H$_2$.  Metal-silicate differentiation may have already begun at this stage.  Stage 3 is the result of chemical equilibration of the silicate and metal melts with the evolved atmospheres. Annotations show the changes in oxygen fugacity and metal density deficits.  Reactions shown are simplifications of the full set, for illustration purposes.  Two or more such embryos combine to form the final Earth.} 
\end{figure}

\end{singlespace}

\vspace{12pt}

 \clearpage
 \section*{Methods}

\noindent{\bf Chemical thermodynamics}

\noindent The linearly independent reactions spanning our reaction space, and strategies for solving the coupled thermodynamic and mass balance equations, were described previously by Schlichting and Young \cite{Schlichting_Young_2022} and are described here briefly.  The reactions comprising our model include speciation reactions in the magma ocean: 
\begin{equation}
{\rm{N}}{{\rm{a}}_{\rm{2}}}{\rm{Si}}{{\rm{O}}_{\rm{3}}} \mathbin{\lower.3ex\hbox{$\buildrel\textstyle\rightarrow\over
{\smash{\leftarrow}\vphantom{_{\vbox to.5ex{\vss}}}}$}} {\rm{N}}{{\rm{a}}_{\rm{2}}}{\rm{O + Si}}{{\rm{O}}_{\rm{2}}};
\end{equation}
\begin{equation}
{\rm{1/2}}\;{\rm{Si}}{{\rm{O}}_{\rm{2}}}{\rm{ +  F}}{{\rm{e}}_{{\rm{metal}}}} \mathbin{\lower.3ex\hbox{$\buildrel\textstyle\rightarrow\over
{\smash{\leftarrow}\vphantom{_{\vbox to.5ex{\vss}}}}$}} {\rm{FeO}}\;{\rm{ + }}\;{\rm{1/2}}\;{\rm{S}}{{\rm{i}}_{{\rm{metal}}}} ;
\label{rxn:R2}
\end{equation}
\begin{equation}
{\rm{MgSi}}{{\rm{O}}_{\rm{3}}} \mathbin{\lower.3ex\hbox{$\buildrel\textstyle\rightarrow\over
{\smash{\leftarrow}\vphantom{_{\vbox to.5ex{\vss}}}}$}} {\rm{MgO  +  Si}}{{\rm{O}}_{\rm{2}}} ;
\end{equation}
\begin{equation}
{{\rm{O}}_{{\rm{metal}}}} + {\rm{ 1/2 S}}{{\rm{i}}_{{\rm{metal}}}} \mathbin{\lower.3ex\hbox{$\buildrel\textstyle\rightarrow\over
{\smash{\leftarrow}\vphantom{_{\vbox to.5ex{\vss}}}}$}} {\rm{ 1/2 Si}}{{\rm{O}}_2} ;
\end{equation}
\begin{equation}
{\rm{2 }}{{\rm{H}}_{{\rm{metal}}}}{\rm{ }} \mathbin{\lower.3ex\hbox{$\buildrel\textstyle\rightarrow\over
{\smash{\leftarrow}\vphantom{_{\vbox to.5ex{\vss}}}}$}} {\rm{ }}{{\rm{H}}_{\rm{2}}}_{,{\rm{silicate}}} ;
\end{equation}
\begin{equation}
{\rm{FeSi}}{{\rm{O}}_{\rm{3}}}{\rm{ }} \mathbin{\lower.3ex\hbox{$\buildrel\textstyle\rightarrow\over
{\smash{\leftarrow}\vphantom{_{\vbox to.5ex{\vss}}}}$}} {\rm{ FeO  +  Si}}{{\rm{O}}_{\rm{2}}} ;
\end{equation}
\begin{equation}
{\rm{2 }}{{\rm{H}}_{\rm{2}}}{{\rm{O}}_{{\rm{silicate}}}}{\rm{  +  S}}{{\rm{i}}_{{\rm{metal}}}}{\rm{ }} \mathbin{\lower.3ex\hbox{$\buildrel\textstyle\rightarrow\over
{\smash{\leftarrow}\vphantom{_{\vbox to.5ex{\vss}}}}$}} {\rm{Si}}{{\rm{O}}_{\rm{2}}}{\rm{  +  2 }}{{\rm{H}}_{{\rm{2, silicate}}}},  
\end{equation}
\noindent reactions that take place in the atmosphere:
\begin{equation}
{\rm{C}}{{\rm{O}}_{{\rm{gas}}}}{\rm{  +  1/2 }}{{\rm{O}}_{{\rm{2, gas}}}} \mathbin{\lower.3ex\hbox{$\buildrel\textstyle\rightarrow\over
{\smash{\leftarrow}\vphantom{_{\vbox to.5ex{\vss}}}}$}} {\rm{ C}}{{\rm{O}}_{{\rm{2, gas}}}}{\rm{ }},
\end{equation}
\begin{equation}
{\rm{C}}{{\rm{H}}_{{\rm{4, gas}}}}{\rm{  +  1/2 }}{{\rm{O}}_{{\rm{2, gas}}}}{\rm{ }} \mathbin{\lower.3ex\hbox{$\buildrel\textstyle\rightarrow\over
{\smash{\leftarrow}\vphantom{_{\vbox to.5ex{\vss}}}}$}} {\rm{ 2 }}{{\rm{H}}_{\rm{2}}}{\rm{  +  CO}},
\end{equation}
\noindent  
\begin{equation}
{{\rm{H}}_{{\rm{2, gas}}}}{\rm{  +  1/2 }}{{\rm{O}}_{{\rm{2, gas}}}}{\rm{ }} \mathbin{\lower.3ex\hbox{$\buildrel\textstyle\rightarrow\over
{\smash{\leftarrow}\vphantom{_{\vbox to.5ex{\vss}}}}$}} {\rm{ }}{{\rm{H}}_{\rm{2}}}{{\rm{O}}_{{\rm{gas}}}}, 
\end{equation}
\noindent and reactions that describe exchange between the atmosphere with the magma ocean:
\begin{equation}
{\rm{FeO }} \mathbin{\lower.3ex\hbox{$\buildrel\textstyle\rightarrow\over
{\smash{\leftarrow}\vphantom{_{\vbox to.5ex{\vss}}}}$}} {\rm{F}}{{\rm{e}}_{{\rm{gas}}}}{\rm{  +  1/2 }}{{\rm{O}}_{{\rm{2,gas}}}},
\end{equation}
\begin{equation}
{\rm{MgO }} \mathbin{\lower.3ex\hbox{$\buildrel\textstyle\rightarrow\over
{\smash{\leftarrow}\vphantom{_{\vbox to.5ex{\vss}}}}$}} {\rm{ M}}{{\rm{g}}_{{\rm{gas}}}}{\rm{ +  1/2 }}{{\rm{O}}_{{\rm{2,gas}}}},
\end{equation}
\begin{equation}
{\rm{Si}}{{\rm{O}}_{\rm{2}}}{\rm{ }} \mathbin{\lower.3ex\hbox{$\buildrel\textstyle\rightarrow\over
{\smash{\leftarrow}\vphantom{_{\vbox to.5ex{\vss}}}}$}} {\rm{ Si}}{{\rm{O}}_{{\rm{gas}}}}{\rm{  +  1/2 }}{{\rm{O}}_{{\rm{2,gas}}}},
\end{equation}
\begin{equation}
{\rm{N}}{{\rm{a}}_{\rm{2}}}{\rm{O }} \mathbin{\lower.3ex\hbox{$\buildrel\textstyle\rightarrow\over
{\smash{\leftarrow}\vphantom{_{\vbox to.5ex{\vss}}}}$}} {\rm{ 2 N}}{{\rm{a}}_{{\rm{gas}}}}{\rm{  +  1/2 }}{{\rm{O}}_{{\rm{2,gas}}}},
\end{equation}
\begin{equation}
{{\rm{H}}_{{\rm{2,gas}}}}{\rm{ }} \mathbin{\lower.3ex\hbox{$\buildrel\textstyle\rightarrow\over
{\smash{\leftarrow}\vphantom{_{\vbox to.5ex{\vss}}}}$}} {\rm{ }}{{\rm{H}}_{{\rm{2,silicate}}}},
\end{equation}
\begin{equation}
{\rm H_2O_ {\rm{gas }}} \mathbin{\lower.3ex\hbox{$\buildrel\textstyle\rightarrow\over
{\smash{\leftarrow}\vphantom{_{\vbox to.5ex{\vss}}}}$}} {\rm H_2O_ {\rm{, silicate}}}
\end{equation}
\begin{equation}
{\rm{C}}{{\rm{O}}_{{\rm{gas}}}}{\rm{ }} \mathbin{\lower.3ex\hbox{$\buildrel\textstyle\rightarrow\over
{\smash{\leftarrow}\vphantom{_{\vbox to.5ex{\vss}}}}$}} {\rm{ C}}{{\rm{O}}_{{\rm{silicate}}}},
\end{equation}
\begin{equation}
{\rm{C}}{{\rm{O}}_{{\rm{2, gas}}}}{\rm{ }} \mathbin{\lower.3ex\hbox{$\buildrel\textstyle\rightarrow\over
{\smash{\leftarrow}\vphantom{_{\vbox to.5ex{\vss}}}}$}} {\rm{ C}}{{\rm{O}}_{{\rm{2, silicate}}}}.
\end{equation}
For each of the 18 reactions among 25 phase components described above there is an equation for the condition for equilibrium.  These equations are of the form  
\begin{equation}
\sum\limits_i {{\nu _i}\ln {x_i} + \left[ {\frac{{\Delta \hat G_{{\rm{rxn}}}^ \circ }}{{RT}} + \sum\limits_g {{\nu _g}\ln (P/{P^ \circ })} } \right]}  = 0.
\label{eqn:f}
\end{equation}
\noindent Here we have used $\mu_i = \Delta \hat G_i^{\rm o} + RT \ln(x_i)$ for the chemical potential of species $i$ corrected for composition where  $\Delta \hat G_i^{\rm o}$ is the Gibbs free energy of formation of $i$ at a standard state of the pure species at temperature and 1 bar pressure, and $R$ is the gas constant.  The sum over index $g$ refers to gas species  in the reaction and index $i$ refers to all species, including the gases.  We  separated the compositional and pressure effects for the gas species, replacing partial pressure (ideal fugacity), for species $g$, by the product $x_g P$ where $P$ is the total gas pressure and $P^{\rm o}$ is the pressure at standard state (chosen as 1 bar here). 

Following Badro et al. \cite{Badro2015}, we replaced mole fractions for Si and O in Equation \ref{eqn:f} with activities $a_i$ where $a_i = \gamma_i x_i$ to acccount for non-ideal competition between O and Si in the metal phase.  The activity coefficient for Si, $\gamma_{\rm Si}$, is given by
\begin{equation}
\begin{aligned}
\ln \gamma_{\rm Si}=&- 6.65\frac{1873}{T}-12.41\frac{1873}{T}\ln(1-x_{\rm Si})
+5\frac{1873}{T} x_{\rm O} \left(1+\frac{\ln(1-x_{\rm O})}{x_{\rm O}}-\frac{1}{(1-x_{\rm Si})}\right)\\
&-5\frac{1873}{T} x_{\rm O}^2\,  x_{\rm Si} 
\left( 
\frac{1}{1-x_{\rm Si}}
+\frac{1}{1-x_{\rm O}}+
\frac{x_{\rm Si}}
{2(1-x_{\rm Si})^2}-1
\right),
\end{aligned}
\end{equation}
\noindent and the activity coefficient for O, $\gamma_{\rm O}$, is obtained using
\begin{equation}
\begin{aligned}
\ln \gamma_{\rm O}=&4.29-\frac{16500}{T}+\frac{16500}{T}\ln(1-x_{\rm O})
+5\frac{1873}{T} x_{\rm Si} \left(1+\frac{\ln(1-x_{\rm Si})}{x_{\rm Si}}-\frac{1}{(1-x_{\rm O})}\right)\\
&-5\frac{1873}{T} x_{\rm Si}^2\, x_{\rm O} 
\left( 
\frac{1}{1-x_{\rm O}}
+\frac{1}{1-x_{\rm Si}}+
\frac{x_{\rm O}}
{2(1-x_{\rm O})^2}-1
\right).
\end{aligned}
\end{equation}
\noindent The activity coefficient for Fe in metal using this approach is near unity (e.g., 0.8 \cite{Corgne2008}). We assume  $\gamma_{\rm Fe}=1$ given the uncertainties in mixing parameters involving H in metal.  The mixing behavior of H in liquid iron metal at high pressures is not as well characterized.  We therefore adopt ideal mixing for H in Fe metal.  Recent results by Li et al. \cite{Li2020_1038} suggest that H pairs mainly with Fe in metal with no preference for bonding with O, consistent with composition-independent, ideal mixing.

Ideal mixing is assumed for silicate melt species. Ideal mixing as applied here has been validated using the MAGMA code \cite{Fegley1987,schaefer2009a} where speciation of simple oxides like FeO and MgO to silicate species like FeSiO$_3$ and  MgSiO$_3$ accounts for first-order activity/composition effects \cite{Hastie1986}. Further discussion of the effects of non-ideal mixing on our results is presented below.  

To these equations we add an additional 7 that account for the summing constraints for each of the 7 elements making up the body:
\begin{equation}
{n_s} - \sum\limits_k {\sum\limits_i {{\eta _{s,i,k}}{x_{i,k}}{N_k}} }  = 0, 
\end{equation}
\noindent where $n_s$ is the moles of element $s$ in the planet, $\eta_{s,i,k}$  is the number of moles of element $s$ in component $i$ of phase $k$, $x_{i,k}$ is the mole fraction of component $i$ in phase $k$, and $N_k$ is the moles of phase $k$ (metal, silicate, or atmosphere). The moles of the phases are treated as variables along with the mole fractions for the species, resulting in 28 variables in the system of equations. The mole fractions for each phase must sum to unity, so our new equations include 
\begin{equation}
\begin{aligned}
1 - {x_{{\rm{MgO}}}} - {x_{{\rm{Si}}{{\rm{O}}_{\rm{2}}}}} - {x_{{\rm{MgSi}}{{\rm{O}}_{\rm{3}}}}} - {x_{{\rm{FeO}}}} - {x_{{\rm{FeSi}}{{\rm{O}}_{\rm{3}}}}} -
{x_{{\rm{N}}{{\rm{a}}_{\rm{2}}}{\rm{O}}}} - \\ {x_{{\rm{N}}{{\rm{a}}_{\rm{2}}}{\rm{Si}}{{\rm{O}}_{\rm{3}}}}} - x_{{{\rm{H}}_{\rm{2}}}}^{{\rm{silicate}}} - 
x_{{{\rm{H}}_{\rm{2}}}{\rm{O}}}^{{\rm{silicate}}} - 
x_{{\rm{CO}}}^{{\rm{silicate}}} - x_{{\rm{C}}{{\rm{O}}_{\rm{2}}}}^{{\rm{silicate}}} = 0
\end{aligned}
\end{equation}
\noindent for the silicate phase,
\begin{equation}
1 - x_{{\rm{Fe}}}^{{\rm{metal}}} - x_{{\rm{Si}}}^{{\rm{metal}}} - x_{\rm{O}}^{{\rm{metal}}} - x_{\rm{H}}^{{\rm{metal}}} = 0
\end{equation}
\noindent for the metal phase, and
\begin{equation}
\begin{aligned}
1 - x_{{\rm{CO}}}^{{\rm{gas}}} - x_{{\rm{C}}{{\rm{O}}_{\rm{2}}}}^{{\rm{gas}}} - x_{{{\rm{O}}_{\rm{2}}}}^{{\rm{gas}}} - x_{{\rm{C}}{{\rm{H}}_{\rm{4}}}}^{{\rm{gas}}} - x_{{{\rm{H}}_{\rm{2}}}}^{{\rm{gas}}} - x_{{{\rm{H}}_{\rm{2}}}{\rm{O}}}^{{\rm{gas}}} - x_{{\rm{Fe}}}^{{\rm{gas}}} - 
x_{{\rm{Mg}}}^{{\rm{gas}}} - x_{{\rm{SiO}}}^{{\rm{gas}}} -x_{{\rm{Na}}}^{{\rm{gas}}} = 0
\end{aligned}
\end{equation}
\noindent for the atmosphere. 

The atmospheric pressure at the surface of the magma ocean, $P_{\rm surface}$ is an additional variable that depends on the mean molecular weight of the atmosphere, and thus must be included in the solutions.  This is accomplished by adding the equation 
\begin{equation}
\left( \frac{P_{\rm surface}}{1 {\rm bar}} \right) = 1.2 \times 10^6 \frac{M_{\rm atm}}{M_{\rm p}} \left( \frac{M_{\rm p}}{M_{\oplus}} \right)^{2/3}
\end{equation}
\noindent to the system of equations to be solved, where the mass of the atmosphere, $M_{\rm atm}$, is obtained from the grams per mole ($\mu$) and the moles of atmosphere, $N_{\rm atm}$.

The 29 non-linear equations in 29 variables were solved using simulated annealing followed by Markov chain Monte Carlo (MCMC) sampling.  We used the Python implementation of simulated annealing  \cite{Xiang_1997} and the  MIT Python implementation of ensemble MCMC \cite{Foreman2019,Goodman2010}.   The combination of simulated annealing followed by MCMC sampling avoids becoming stranded in local minima.  We used the thermodynamic data, including the treatment of H$_2$ solubility,  described in detail by Schlichting and Young \cite{Schlichting_Young_2022} with the exception that we replaced the free energy of reaction for reaction 2 (Equation \ref{rxn:R2}) with that given by Corgne et al. \cite{Corgne2008}. The use of free energies for the reactions permits application of self-consistent free energies of formation for the various species in our reaction network, and supplants distribution coefficients for partitioning of elements between phases that are not necessarily internally consistent.  

\noindent{\bf Embryo primary atmospheres}

\noindent Extended Data Figure 1 shows the relationship between the mass of planetary embryos, their surface temperatures, and the potential coexistence of a surface magma ocean and hydrogen-rich primary atmospheres following Ginzburg et al. \cite{ginzburg2016a}.  The boundary between solid and liquid silicate at the condensed surface is taken to be 1500 K in the figure.  Mars is shown to be sufficiently massive to have had a primary atmosphere, but not massive enough to have this atmosphere in coexistence with a surface magma ocean. The crossing of the two blue lines in the figure defines the minimum mass for an embryo to have simultaneously a primary hydrogen-rich atmosphere and a surface magma ocean.

 \noindent{\bf Silicon in metal}
 
 \noindent Extended Data Figure 2 shows our results for the weight percent of Si in metal as a function of core-mantle equilibration temperature and initial mass fraction of the H$_2$-rich primary atmosphere.  This figure is comparable to the panels shown in Figure 1  of the main text. 

\noindent{\bf Sensitivity tests}

\noindent We examined the effects of the assumption of ideal mixing between  H$_2$O and silicate and between H$_2$ and silicate on our results.  We also evaluated the effects of non-ideal mixing of H in the Fe metal alloy phase.  Our fiducial model recalculated to include all of these potential non-ideal mixing behaviors is shown in Supplementary Data Table 4 and can be compared directly with the results given in Supplementary Data Table 1. Details of the non-ideal mixing models are provided in this section. 

Kova{\v c}evi{\'c} et al.\ \cite{Kovacevic_2022} reported complete miscibility between H$_2$O and MgSiO$_3$ at temperatures down to 4500 K at 40 GPa, and at higher temperatures at higher pressures.  Extrapolation to the $\sim 100$ bar surface pressures in our models suggests a consolute temperature, $T_{\rm c}$, of $\sim 4000$ K  by analogy with supercritical silicate liquid and vapor \cite{Xiao2018}. Assuming a symmetrical (regular) mixing model based on this observation, an activity coefficient for H$_2$O in melt can be obtained from $\ln{\gamma_{\rm H_2O}^{\rm silicate}}=(1-x_{\rm H_2O}^{\rm silicate})^2 W/(RT)$ where the interaction parameter $W$ is given by $W=2 R T_{\rm c}$ and $T_{\rm c}$ is the temperature at the crest of the silicate-H$_2$ solvus. The interaction parameter $W$ for $T_c = 4000$ K is  $66512$ J/mole.  

The activity/composition relationship for H$_2$ in silicate melt is not well characterized.  However, we can speculate that non-ideal mixing could be evidenced by complete miscibility between H$_2$ vapor and silicate melt at sufficiently high temperatures.  This speculation is supported by as yet unpublished {\it ab initio} calculations at UCLA. Based on these preliminary calculations,  we consider a value for the consolute temperature of 4500 K for pressures of interest, leading to $W = 74829$ J/mole.  

For H in the liquid iron alloy we used $\epsilon$ interaction parameters.  The $\epsilon$ notation for interaction parameters is commonly used for alloys, representing a Taylor series expansion for the log of the activity coefficient of interest.  In the case of a dilute solute $i$ interaction with species $j$ we can relate $\epsilon_i^j$ to binary interaction parameters $W$ using $\epsilon_i^j=-2 W/(RT)$. Boorstein et al.\ \cite{Boorstein1974} found values for $\epsilon_{\rm H}^{\rm Si}$ in liquid Fe of $3.5$ (a ternary interacton parameter) and Waseda  \cite{Waseda_2012} reports that $\epsilon_{\rm H}^{\rm O}$ is similar to $\epsilon_{\rm H}^{\rm Si}$.  We therefore applied a pseudo-ternary mixing model for H in molten Fe alloy, following Righter et al.\ \cite{Righter_2020}, such that

\begin{align}
\ln{\gamma _{\rm{H}}} &=  - \epsilon _{\rm{H}}^{{\rm{Si}}}{x'_{{\rm{Si}}}}\left( {1 + \frac{{\ln (1 - {x'_{{\rm{Si}}}})}}{{{x'_{{\rm{Si}}}}}} - \frac{1}{{1 - {x_{\rm{H}}}}}} \right)\nonumber \\
 &+ \epsilon _{\rm{H}}^{{\rm{Si}}}(x'_{{\rm{Si}}})^2 {x_{\rm{H}}}\left( {\frac{1}{{(1 - {x_{\rm{H}}})}} + \frac{1}{{(1 - {x'_{{\rm{Si}}}})}} + \frac{{{x_H}}}{{2{{(1 - {x_{\rm{H}}})}^2}}} + 1} \right),
\end{align}
where $x_{\rm H}$ is the mole fraction of H in the metal alloy and  $x'_{\rm Si}$ is the sum of the mole fractions of Si and O in the metal. 

The overall effect of the three non-ideal mixing solution models described here for H$_2$ and H$_2$O in silicate and H in Fe metal alloy is a change in the H$_2$O/H$_2$ ratio in the silicate melt and a doubling of the mass fraction of the atmosphere with no significant change in the water concentration in the atmosphere (Supplementary Data Table 4). The density deficit in the metal core changes from $8.0\%$ to $8.2\%$, a change that would be easily accommodated in our models by slight adjustments to the initial mass fraction of the primary atmosphere. The oxygen fugacity of the mantle is essentially unchanged.  We find, therefore, that the effect of inclusion of non-ideal mixing of H$_2$O and H$_2$ in silicate melt, and non-ideal mixng of H in Fe alloy, is to increase the efficacy of water production with little change to the other salient features of our results.  The reason for the robust nature of our solution is that the concentrations of water and hydrogen in the silicate are able to compensate for the changes in thermodynamic activities by virtue of the lever-rule effect; most H exists in the metal phase with the remainder partitioned between the atmosphere, as water, and the silicate melt. The changes in silicate H$_2$O and H$_2$ concentrations are due primarily to the activity coefficients for these species, with non-ideal mixing of H in the metal having negligible impact on the results. 

As an additional sensitivity test, we considered the effects of altering the precise values for H partitioning between silicate and metal.  We do this because while the free energies of formation of H$_2$ in silicate and H in metal were derived in a self-consistent manner, as described below, these values are associated with unknown uncertainties.  For this purpose we reran our calculations  with the logarithm of the equilibrium constant for reaction 5, $\ln{K_{\rm eq,R5 }} = -\Delta \hat G_{\rm rxn, R5}^{\rm o}/(RT)$, multiplied by factors of $1.5$ and $0.5$, respectively. 

The standard-state free energy of reaction 5 \cite{Schlichting_Young_2022} used here is obtained from $\Delta \hat G_{\rm rxn, R5}^{\rm o}= \Delta \hat G_{\rm H_2}^{\rm o,silicate}-2\Delta \hat G_{\rm H}^{\rm o,metal}$ where $\Delta \hat G_{\rm H_2}^{\rm o,silicate}$  at $T$ and 1 bar was obtained from the free energy of the reaction H$_{2, { \rm gas}}$ = H$_{2, {\rm melt}}$  after Hirschmann et al.\ \cite{Hirschmann2012}, and $\Delta \hat G_{\rm H_2}^{\rm o}$ for gas from NIST.  The $\Delta \hat G_{\rm H}^{\rm o}$ for metal was obtained from  the reaction Fe + H$_2$O$_{\rm melt}$ = FeO + 2H  by regression of $\ln{K_{\rm eq}}$ vs. $1/T$ reported by Okuchi\cite{Okuchi1997}, yielding $\Delta \hat G_{\rm rxn, Okuchi97}^{\rm o} = 143589.7-69.1 T$ (J/mole),  $\Delta \hat G_{\rm H_2O}^{\rm o}$ for silicate melt using the solubility data of Moore et al. \cite{Moore1998} for the reaction H$_2$O$_{\rm gas}$ = H$_2$O$_{\rm melt}$\cite{Schlichting_Young_2022}, and the standard-state  free energies of formation for H$_2$O gas, liquid FeO, and Fe from NIST. From these values, $\Delta \hat G_{\rm H}^{\rm o,metal} = 1/2( \Delta \hat G_{\rm rxn, Okuchi97}^{\rm o}- \Delta \hat G_{\rm FeO}^{\rm o} +\Delta \hat G_{\rm Fe}^{\rm o}+\Delta \hat G_{\rm H_2O}^{\rm o,silicate})$.  

Our results with these changes in the logarithm of the equilibrium constant describing partitioning of H between silicate and metal are negligible, with the largest change being a decrease in the mass of the atmosphere in the case of increasing $\Delta \hat G_{\rm rxn, R5}^{\rm o}$ by $1.5$.  In the latter case, the result in Supplementary Data Table 4 is modified by lowering the pressure of the atmosphere at the magma ocean surface from $\sim 300$ bar to $\sim 200$ bar with no change in the water-rich composition of the atmosphere and no change in the density deficit of the metal core.  In the case of decreasing $\Delta \hat G_{\rm rxn, R5}^{\rm o}$ by a factor of $2\times$, results were virtually unchanged. These results can also be understood as the manifestation of the lever rule in which the large amount of hydrogen in the metal phase relative to the total mass of hydrogen in the other reservoirs dominates the behavior.  To see this, we can describe the activity of H in metal, $a_{\rm H}$, as a product of the equilibrium constants for reactions 5 and 15, arriving at the expression $a_{\rm H}=\sqrt{ K_{\rm eq, R5}K_{\rm eq, R15} f_{\rm H_2} }$ where $f_{\rm H_2}$ is the fugacity of H$_2$ in the atmosphere.  In the case of the embryos, the activity of H in the metal alloy exerts influence over the partial pressure (fugacity) of H$_2$ in the atmosphere due to the high mass of hydrogen relative to that in the atmosphere.

 \noindent{\bf Pressure vs radius for embryos}
 
\noindent The pressure vs.\ radius relationships described in the text are presented in Extended Data Figure 3. The pressure-radius relations are obtained by solving for the mass, $m(r)$, contained within a given radius, $r$ \cite{seager2007a}
\begin{equation}\label{emr1}
    \frac{d m(r)}{d r}=4\pi r^2 \rho(r)
\end{equation}
together with the requirement for hydrostatic equilibrium 
\begin{equation}\label{emr2}
    \frac{dP(r)}{dr}=-\frac{G m(r) \rho(r)}{r^2},
\end{equation}
where 
\begin{equation}
   P(r)=f(\rho(r),T(r)),
\end{equation}
and $f$ represents the appropriate equation of state (EOS). $P(r)$, $\rho(r)$, and $T(r)$ are the radially-dependent pressure, density and temperature of the body, and $G$ is the gravitational constant.

We integrate equations (\ref{emr1}) and (\ref{emr2}) from the embryo's center, using the inner boundary condition $m(0)=0$ and $P(0)=P_{\rm center}$. The outer boundary condition is given by $P(R_{\rm p})=0$.  We switch from one EOS to the other while maintaining continuity in pressure across the core-mantle boundary. The embryo mass, $M_{\rm p}$ and radius, $R_{\rm p}$, are uniquely determined by $P(0)=P_{\rm center}$ and $P(R_{\rm p})=0$ for the Earth-like mass fractions for the metal core used in our models. 

Extrapolation of the data of Kuwayama et al.  \cite{Kuwayama2020} yields an uncompressed density for liquid Fe metal, $\rho_0$, of $7.2 \pm 0.1$ g/cm$^3$. We used this uncompressed density for liquid Fe with a Vinet EOS \cite{seager2007a,Anderson2001}.  For the silicate we use the enstatite (MgSiO$_3$) third-order Birch-Murnagham EOS of Karki et al.\ \cite{Karki2000} described by Seager et al.\ \cite{seager2007a} as a proxy for silicate melt.

Adiabatic temperature gradients for the embryo mantles were calculated using 
\begin{equation}
   \left(\frac{dT}{dr}\right)_{S}=\frac{\alpha g T}{c_{P}}
\end{equation}
where $\alpha$ is the expansivity for MgSiO$_3$, $c_{P}$ is the isobaric specific heat for the melt, and $g$ is the gravitational acceleration.  We used $\alpha=2.69\times 10^{-5}+2.13\times 10^{-8}(T-300)$ from Katsura et al. \cite{Katsura2009} with a pressure correction factor of $1.04$ for the mantle as a whole.  Temperature-dependent $c_{P}$ for liquid MgSiO$_3$ was obtained from the NIST thermodynamic database.  Numerical integrations with $r$ yield a general relationship between potential temperature and core-mantle temperature such that $T_{\Theta}/T_{\rm core-mantle} \sim 0.78$.

\noindent {\bf Isotope fractionation}

\noindent We investigated the implications of our model for Fe, Si, and H stable isotope ratios, as these ratios are often used to constrain the differentiation history of Earth \cite{Shahar2020,Young2015} and the origin of Earth's water \cite{Hallis2015}.  

We find that the embryo partitioning model explains the offset in $^{57}$Fe/$^{54}$Fe (expressed as $\delta^{57}{\rm Fe}$ in per mil) between chondrites of all varieties and the bulk silicate Earth.  Recent estimates suggest that bulk silicate Earth has a $\delta^{57}{\rm Fe} = 0.05 \pm 0.01$ value relative to chondrites \cite{Sossi2016}.  This offset between Earth and its presumed starting materials is reproduced by our model shift in bulk silicate $\delta^{57}{\rm Fe}$ of $0.057$ \textperthousand\, relative to the bulk starting material (Extended Data Figure 4).   
 
 We used the prescription for $^{57}$Fe/$^{54}$Fe fractionation as a function of temperature and pressure determined experimentally by Ni and Shahar et al. \cite{Ni2022}. The reduced partition function ratio, $\beta$, for the silicate is given by
 \begin{equation}
10^3 \ln(\beta_{\rm Fe-silicate})=\frac{4281.4 (4.65 P({\rm GPa}) + 150.4)}{T^2}.
\end{equation}
 The value for FeH alloy is taken to be
 \begin{equation}
10^3 \ln(\beta_{\rm FeH})=\frac{4281.4 (2.597 P({\rm GPa}) + 119.017)}{T^2}.
\end{equation}
 The fractionation factor, expressed as $10^3\ln(\alpha_{\rm silicate-FeH})$ $\sim \delta^{57}{\rm Fe}_{\rm silicate} - \delta^{57}{\rm Fe}_{\rm FeH}$ is then
\begin{equation}
10^3 \ln(\alpha_{\rm silicate-FeH})=10^3 \ln(\beta_{\rm Fe-silicate})-10^3 \ln(\beta_{\rm FeH}).
\end{equation} 
 We used the FeH alloy for the entire metal since the mole fractions of H in our Fe metal phases are high ($\sim 0.27$), and the effect of different Fe/H ratios on the fractionation factors are not yet understood. \

In order to evaluate the Si isotope effects, we used the $^{30}$Si/$^{28}$Si fractionation determined both by experiments and by analysis of equilibrated aubrites \cite{Young2015}, yielding  
 \begin{equation}
10^3 \ln(\alpha_{\rm silicate-metal})=\frac{7.64\times 10^6}{T^2}.
\end{equation}

Similar to the Fe isotope effects described in the main text, our model predicts that the $^{30}$Si/$^{28}$Si ratio of the bulk silicates (expressed as $\delta^{30}$Si) should be greater than the source material by $0.06$ \textperthousand.  This offset is smaller than the observed difference between Earth and many chondrites of about $0.2$ \textperthousand \cite{Pringle2014} and is much smaller than the difference between Earth and E chondrites \cite{Savage2012}, despite the similarity between Earth and E chondrites in virtually every other isotope system.  Our results do not alleviate some of the vexing aspects of Si isotope variability in the Solar System. In general, the apparently anomalous Si isotopic composition of E chondrites may be related to the apparent excess in Si in these rocks ($\sim 20\%$ atomic) compared with solar refractory metal abundances.  

The D/H of the Sun (D/H $= 2.0 \times 10^{-5}$), and thus the protosolar gas, expressed as $\delta$D relative to ocean water (D/H $= 1.558 \times 10^{-4}$), is $\sim -865$ \textperthousand \cite{Geiss2003}. Recent estimates for the bulk D/H ratio of Earth's hydrogen yield a value for $\delta$D relative to ocean water of $-218$ \textperthousand \cite{Hallis2015}.  It is conventional to ascribe the origin of Earth's water to late addition of small bodies, asteroids or comets, because virtually all sources of D/H in the Solar System other than protosolar gas, including comets \cite{Altwegg2015,BockeleeMorvan2012}, chondrite meteorites \cite{Alexander2012}, and water in icy moons \cite{Waite2009}, have $\delta$D $\ge 0$, and often much greater than zero (Extended Data Figure 5). Calculations based on mixing of asteroids and/or comets with a solar gas based on D/H ratios  suggest that there may be a solar gas component comprising roughly $25$ to $40 \%$ of hydrogen deep within the Earth, depending on whether the non-solar component is chondrite-like or comet-like. For comparison, $^{20}$Ne/$^{22}$Ne data suggest Earth acquired $\sim$2/3 of its neon from the solar gas and the remaining 1/3 from E chondrite-like material \cite{Lammer2020}.

In general, any reactions between H$_2$ and H$_2$O, either at equilibrium or kinetically controlled through reactive intermediates, result in D/H of water being greater than that of H$_2$, so the use of D/H as a tracer involves numerous degeneracies.  At room temperature and below, the kinetic or equilibrium preferential transfer of D to water from solar-like dihydrogen will result in D/H in water comparable to or greater than those observed in Solar System water. Therefore, it is likely that a variety of different processes were at work to cause the generally high D/H of water in the Solar System.

Motivated by the model presented here, we consider the possibility that the high D/H of Earth's water and deep interior could be the residual effect of low-temperature exchange of D and H between H$_2$O and H$_2$ in embryo atmospheres, and the sequestration of large fractions of H in the metallic cores. The convecting silicate magma oceans would serve as the conveyor linking  the atmospheres and metal cores in this context. Hydrogen isotope exchange between radical derivatives of dihydrogen gas and water at temperatures similar to the equilibrium temperature at the top of an embryo atmosphere at about 1 AU (255 K present day, 237 K accounting for a fainter Sun $4.6$ Gyr before present\cite{Gough1980}) will result in high D/H in the atmosphere \cite{Genda2008} if there is a balancing sink for the low D/H complement of hydrogen.  In the model of Genda and Ikoma\cite{Genda2008}, higher D/H in atmospheric water is balanced by slow (over billion year timescales) hydrodynamic escape of isotopically light hydrogen\cite{Lammer2014}.  Sharp and Olsen \cite{Sharp2022} also advanced the idea that the high D/H of Earth's water was due to hydrodynamic escape, in this case as H rather than H$_2$.  In our model, the requisite low-D/H sinks for hydrogen that balance the high D/H of terrestrial water are the metal cores; hydrogen isotope equilibration  between embryo metal cores, silicate mantles, and atmospheres results in a high D/H atmospheres and low D/H metal cores.  

This explanation relies on  a faster rate of isotope exchange high in the atmosphere where temperatures are low ($< 300$ K) than exchange at high temperatures near the surface of the magma ocean.  A detailed atmospheric model is beyond the scope of the present study, but we note that D/H exchange is expected to be orders of magnitude faster in the upper atmosphere where kinetics mediated by radicals result from low densities and photochemistry compared with exchange between H$_2$O and H$_2$ molecules at high temperatures and densities. For example, the forward rate constant for the reaction HD + OH $\rightarrow$ HDO + H at a temperature of 237 K (representing the 1 AU equilibrium temperature for the stratosphere of an embryo $4.5$ Gyr before present) is $1.9\times 10^{-15}$ cm$^3$ s$^{-1}$ \cite{Yang2013} . This reaction represents transfer of D from hydrogen to OH, and ultimately water, in embryo stratospheres. The rate constant for the reaction H$_2$ + HDO $\rightarrow$ HD + H$_2$O is $1.4\times 10^{-23}$ cm$^3$ s$^{-1}$ at 1973 K, the latter temperature being an altitude-averaged temperature for the adiabatic tropospheres of atmospheres implied by our models. This represents return of D from water to hydrogen in the tropospheres.  While the details rely on a number of uncertain parameters, including ratios of e-folding times for exchange and residence times in the various reservoirs, it is unlikely that this disparity in rate constants can be overcome by the relative masses of H$_2$O and H$_2$ in the embryo stratospheres and tropospheres (for reference, Earth's ratio of moles in the stratosphere to moles in the troposphere is $0.1$).\cite{Holton1990}  Vigorous mixing of water and hydrogen between the stratosphere and troposphere should therefore result in D/H isotope fractionation dominated by the lower temperatures.  The turnover time for Earth's stratosphere is about $2.5$ yr due to transfer back and forth between the stratosphere and troposphere \cite{Holton1990}. We speculate that it is likely that H$_2$ and the underlying magma oceans could have reached isotopic equilibrium during the earliest phases of dissolving molecular hydrogen into the melts \cite{Pahlevan2019}. 

As a plausibility exercise based on the reasoning described above, we calculated the relative abundances of the deuterated species in the stratospheres and tropospheres from the surface densities (column densities) of an adiabatic troposphere at the surface of the magma ocean and the surface density of the stratosphere as measured at the tropopause using a simplified atmosphere \cite{Young2019}. The mixing ratio of H$_2$O was taken to be $0.8$ with the remainder of the atmosphere being H$_2$. 
The temperature at the base of the troposphere was taken to be the potential temperature of the magma oceans in our model (2350 K), the height of the tropopause was assumed to be where the temperature reaches the equilibrium temperature, and the stratosphere was taken to be isothermal, extending 5 surface scale heights. From this we obtain a ratio of moles of HD in the stratosphere to moles of HDO in the troposphere, $N_{\rm HD}$/$N_{\rm HDO}$, of $3\times 10^{-4}$. Converting the rate constants for the two reactions described above to $k$(moles$^{-1}$ yr$^{-1}$) using $k$(cm$^3$ s$^{-1}$)$L/V$ where $L$ is Avagadro's number and $V$ are the similar volumes of each of the two reservoirs, one obtains rate constants of 
$4.3\times 10^{-10}$ moles$^{-1}$ yr$^{-1}$ for HD + OH $\rightarrow$ HDO + H at a temperature of 237 K and $1.1\times 10^{-18}$ moles$^{-1}$ yr$^{-1}$ for
H$_2$ + HDO $\rightarrow$ HD + H$_2$O at 1973 K.
From these values one obtains a time constant for D/H exchange in the stratosphere of $5\times 10^{-7}$ yr for an OH/H$_2$O ratio of $\sim 10^{-4}$ and a time constant for D/H exchange in the troposphere of $1\times 10^{-4}$ yr. Smaller concentrations of OH result in proportionally larger time constants in the stratosphere; equal time constants for exchange in the stratosphere and troposphere would require OH/H$_2$O ratios of about $10^{-7}$.

We take the shorter time constant for D/H exchange in the stratosphere relative to the troposphere to mean that a low-temperature for H$_2$O-H$_2$ D/H exchange in the atmosphere on average is plausible.  Equilibrium D/H exchange between H$_2$O and H$_2$ has been inferred for the martian atmosphere at similar temperatures of $\sim 200$ K \cite{Krasnopolsky1998}.  We note for comparison that various terrestrial analogs exist for isotopic anomalies formed in the stratosphere to persist in the troposphere, including in the cases of oxygen and nitrogen isotopologues \cite{Young2014,Yeung2017}.   

Here we adopted a low temperature of exchange between atmospheric H$_2$O and H$_2$ and calculated the equilibrium $\delta$D values associated with our model relative to the primordial hydrogen atmospheres of embryos.  Our calculations assume hydrogen isotopic equilibrium between atmospheres, magma oceans, and molten metal cores. We obtained $\delta$D values for atmospheric H$_2$, atmospheric H$_2$O, H in silicate melt, and H in metal at equilibrium by solving the hydrogen isotope mass balance equation $\delta$D (bulk) = $\sum_i x_{{\rm H},i}\, \delta{\rm D}_i$, where $x_{{\rm H},i}$ refers to the hydrogen fraction for each phase $i$, together with the equilibrium differences between reservoirs $i$ and $j$ of form $\delta$D$_i - \delta$D$_j = 10^3 \ln(\alpha_{i-j})$ for the high-$T$ fractionations, and $\delta$D$_{\rm H_2O}$ $-$ $\alpha_{\rm H_2O-H_2}\delta$D$_{\rm H_2} = 10^3(\alpha_{\rm H_2O - H_2}-1)$ for the lower-$T$, and thus much larger, atmospheric H$_2$O and H$_2$ fractionation.  We thus have four equations for four unknown $\delta$D values given three known fractionation factors, $\alpha$, and an assigned value for  $\delta$D (system) (taken to be  $0$ for reference).  The equations to be solved, in matrix form, are

\begin{equation}
 \quad \left[ {\begin{array}{*{20}{c}}
x_{\rm H,H_2} & x_{\rm H,H_2O} & x_{\rm H,sil} & x_{\rm H,metal} \\
0&0&1&-1\\
-\alpha_{\rm H_2O-H_2}&1&0&0\\
-1&0&1&0\\
\end{array}} \right]\quad \left[ \begin{array}{l}
\delta{\rm D_{H_2}} \\
\delta{\rm D_{H_2O}}\\
\delta{\rm D_{sil}}\\
\delta{\rm D_{metal}}
\end{array} \right]=
\left[ \begin{array}{l}
\delta{\rm D_{system}}=0\\
10^3 \ln(\alpha_{\rm slicate-metal})\\
10^3 (\alpha_{\rm H_2O - H_2}-1)\\
 10^3 \ln(\alpha_{\rm silicate - H_2}) \\
\end{array} \right]\quad  
\end{equation}

We used fractionation factors described previously by Young et al. \cite{Young2015}.  The fractionation factors for D/H, expressed as per mil differences in $\delta$D values, or $10^3\ln(\alpha)$, include the value for silicate and metal (i.e., $\delta$D silicate $-$ $\delta$D metal):
 
 \begin{equation}
 10^3 \ln(\alpha_{\rm slicate-metal}) = \frac{4.5\times 10^8}{T_{\rm core-mantle}^2},
 \end{equation}
 the value for H$_2$O and H$_2$:
 \begin{equation}
 10^3 \ln(\alpha_{\rm H_2O - H_2}) = \frac{215.8\times 10^3}{T_{\rm atmosphere}}+91.769\left(\frac{10^3}{T_{\rm atmosphere}}\right)^2-11.419\left(\frac{10^3}{T_{\rm atmosphere}}\right)^3 -87.44,
 \end{equation}
 and  the value between silicate and H$_2$ gas based on the difference between $10^3 \ln(\alpha_{\rm slicate-metal})$  and the fractionation between H$_2$ and metal \cite{Young2015}:
 \begin{equation}
 10^3 \ln(\alpha_{\rm silicate - H_2}) = \frac{4.5\times 10^8}{T_{\rm surface}^2}-\frac{3.261\times 10^8}{T_{\rm surface}^2}.
 \end{equation}
 Each equilibrium is specified to occur at the core-mantle boundary, in the atmosphere, or at the atmosphere-magma ocean interface at the specified temperatures of 3000 K, 237 K, and 2350 K, respectively.  Results are shown in Extended Data Figure 5. If the structure of the atmospheres of the embryos was such that D/H exchange in the atmosphere was dominated by lower temperatures, the D/H of terretrial water is explained by our model.  The implied D/H of the primordial atmosphere source is similar to that for Uranus and Neptune and somewhat greater than the solar value (Extended Data Figure 5), allowing for some late additions of water to Earth by high-D/H bolides.  We emphasize that this model is possible because of the metal cores serving as low-D/H reservoirs.

\noindent {\bfseries Data Availability} The python code used for models shown in Figure 1 in this study is available at GitHub. Example model results are available as Supplementary information tables.

\begin{singlespace}
\bibliographystyle{naturemag}

\begin{thebibliography}{10}
\expandafter\ifx\csname url\endcsname\relax
  \def\url#1{\texttt{#1}}\fi
\expandafter\ifx\csname urlprefix\endcsname\relax\def\urlprefix{URL }\fi
\providecommand{\bibinfo}[2]{#2}
\providecommand{\eprint}[2][]{\url{#2}}

\bibitem{bean2021}
\bibinfo{author}{{Bean}, J.~L.}, \bibinfo{author}{{Raymond}, S.~N.} \&
  \bibinfo{author}{{Owen}, J.~E.}
\newblock \bibinfo{title}{{The Nature and Origins of Sub-Neptune Size
  Planets}}.
\newblock \emph{\bibinfo{journal}{Journal of Geophysical Research (Planets)}}
  \textbf{\bibinfo{volume}{126}}, \bibinfo{pages}{e06639}
  (\bibinfo{year}{2021}).

\bibitem{Wetherill1978}
\bibinfo{author}{{Wetherill}, G.~W.}
\newblock \bibinfo{title}{{Accumulation of the Terrestrial Planets}}.
\newblock In \bibinfo{editor}{{Gehrels}, T.} \& \bibinfo{editor}{{Matthews},
  M.~S.} (eds.) \emph{\bibinfo{booktitle}{IAU Colloq. 52: Protostars and
  Planets}}, \bibinfo{pages}{565} (\bibinfo{year}{1978}).

\bibitem{Rubie2015}
\bibinfo{author}{Rubie, D.} \emph{et~al.}
\newblock \bibinfo{title}{Accretion and differentiation of the terrestrial
  planets with implications for the compositions of early-formed solar system
  bodies and accretion of water}.
\newblock \emph{\bibinfo{journal}{Icarus}} \textbf{\bibinfo{volume}{248}},
  \bibinfo{pages}{89--108} (\bibinfo{year}{2015}).

\bibitem{Albarede2009}
\bibinfo{author}{Albar{\`e}de, F.}
\newblock \bibinfo{title}{Volatile accretion history of the terrestrial planets
  and dynamic implications}.
\newblock \emph{\bibinfo{journal}{Nature}} \textbf{\bibinfo{volume}{461}},
  \bibinfo{pages}{1227--1233} (\bibinfo{year}{2009}).

\bibitem{Cartier2019}
\bibinfo{author}{Cartier, C.} \& \bibinfo{author}{Wood, B.~J.}
\newblock \bibinfo{title}{{The Role of Reducing Conditions in Building
  Mercury}}.
\newblock \emph{\bibinfo{journal}{Elements}} \textbf{\bibinfo{volume}{15}},
  \bibinfo{pages}{39--45} (\bibinfo{year}{2019}).

\bibitem{Dauphas2017}
\bibinfo{author}{Dauphas, N.}
\newblock \bibinfo{title}{The isotopic nature of the earth's accreting material
  through time}.
\newblock \emph{\bibinfo{journal}{Nature}} \textbf{\bibinfo{volume}{541}},
  \bibinfo{pages}{521--524} (\bibinfo{year}{2017}).

\bibitem{Sikdar2020}
\bibinfo{author}{Sikdar, J.} \& \bibinfo{author}{Rai, V.~K.}
\newblock \bibinfo{title}{Si-mg isotopes in enstatite chondrites and accretion
  of reduced planetary bodies}.
\newblock \emph{\bibinfo{journal}{Scientific Reports}}
  \textbf{\bibinfo{volume}{10}}, \bibinfo{pages}{1273} (\bibinfo{year}{2020}).

\bibitem{Javoy1995}
\bibinfo{author}{Javoy, M.}
\newblock \bibinfo{title}{The integral enstatite chondrite model of the earth}.
\newblock \emph{\bibinfo{journal}{Geophysical Research Letters}}
  \textbf{\bibinfo{volume}{22}}, \bibinfo{pages}{2219--2222}
  (\bibinfo{year}{1995}).

\bibitem{Nittler2011}
\bibinfo{author}{Nittler, L.~R.} \emph{et~al.}
\newblock \bibinfo{title}{The major-element composition of mercury\&\#x2019;s
  surface from messenger x-ray spectrometry}.
\newblock \emph{\bibinfo{journal}{Science}} \textbf{\bibinfo{volume}{333}},
  \bibinfo{pages}{1847--1850} (\bibinfo{year}{2011}).

\bibitem{Shepard2015}
\bibinfo{author}{Shepard, M.~K.} \emph{et~al.}
\newblock \bibinfo{title}{A radar survey of m- and x-class asteroids. iii.
  insights into their composition, hydration state, \& structure}.
\newblock \emph{\bibinfo{journal}{Icarus}} \textbf{\bibinfo{volume}{245}},
  \bibinfo{pages}{38--55} (\bibinfo{year}{2015}).

\bibitem{Zellner1977}
\bibinfo{author}{Zellner, B.}, \bibinfo{author}{Leake, M.},
  \bibinfo{author}{Morrison, D.} \& \bibinfo{author}{Williams, J.}
\newblock \bibinfo{title}{The e asteroids and the origin of the enstatite
  achondrites}.
\newblock \emph{\bibinfo{journal}{Geochimica et Cosmochimica Acta}}
  \textbf{\bibinfo{volume}{41}}, \bibinfo{pages}{1759--1767}
  (\bibinfo{year}{1977}).

\bibitem{Piani2020}
\bibinfo{author}{Piani, L.} \emph{et~al.}
\newblock \bibinfo{title}{Earth\&\#x2019;s water may have been inherited from
  material similar to enstatite chondrite meteorites}.
\newblock \emph{\bibinfo{journal}{Science}} \textbf{\bibinfo{volume}{369}},
  \bibinfo{pages}{1110--1113} (\bibinfo{year}{2020}).

\bibitem{Petigura2013b}
\bibinfo{author}{Petigura, E.~A.}, \bibinfo{author}{Howard, A.~W.} \&
  \bibinfo{author}{Marcy, G.~W.}
\newblock \bibinfo{title}{Prevalence of earth-size planets orbiting sun-like
  stars}.
\newblock \emph{\bibinfo{journal}{Proceedings of the National Academy of
  Sciences}} \textbf{\bibinfo{volume}{110}}, \bibinfo{pages}{19273--19278}
  (\bibinfo{year}{2013}).

\bibitem{weiss2014a}
\bibinfo{author}{{Weiss}, L.~M.} \& \bibinfo{author}{{Marcy}, G.~W.}
\newblock \bibinfo{title}{{The Mass-Radius Relation for 65 Exoplanets Smaller
  than 4 Earth Radii}}.
\newblock \emph{\bibinfo{journal}{The Astrophysical Journal}}
  \textbf{\bibinfo{volume}{783}}, \bibinfo{pages}{L6} (\bibinfo{year}{2014}).

\bibitem{fulton2017a}
\bibinfo{author}{{Fulton}, B.~J.} \emph{et~al.}
\newblock \bibinfo{title}{{The California-Kepler Survey. III. A Gap in the
  Radius Distribution of Small Planets}}.
\newblock \emph{\bibinfo{journal}{Astronomical Journal}} \textbf{\bibinfo{volume}{154}},
  \bibinfo{pages}{109} (\bibinfo{year}{2017}).

\bibitem{Berger2020}
\bibinfo{author}{{Berger}, T.~A.}, \bibinfo{author}{{Huber}, D.},
  \bibinfo{author}{{Gaidos}, E.}, \bibinfo{author}{{van Saders}, J.~L.} \&
  \bibinfo{author}{{Weiss}, L.~M.}
\newblock \bibinfo{title}{{The Gaia-Kepler Stellar Properties Catalog. II.
  Planet Radius Demographics as a Function of Stellar Mass and Age}}.
\newblock \emph{\bibinfo{journal}{The Astronomical Journal}}
  \textbf{\bibinfo{volume}{160}}, \bibinfo{pages}{108} (\bibinfo{year}{2020}).

\bibitem{owen2013a}
\bibinfo{author}{{Owen}, J.~E.} \& \bibinfo{author}{{Wu}, Y.}
\newblock \bibinfo{title}{{Kepler Planets: A Tale of Evaporation}}.
\newblock \emph{\bibinfo{journal}{The Astrophysical Journal}}
  \textbf{\bibinfo{volume}{775}}, \bibinfo{pages}{105} (\bibinfo{year}{2013}).

\bibitem{gupta2019a}
\bibinfo{author}{{Gupta}, A.} \& \bibinfo{author}{{Schlichting}, H.~E.}
\newblock \bibinfo{title}{{Sculpting the valley in the radius distribution of
  small exoplanets as a by-product of planet formation: the core-powered
  mass-loss mechanism}}.
\newblock \emph{\bibinfo{journal}{Monthly Notices of the Royal Astronomical Society}} \textbf{\bibinfo{volume}{487}},
  \bibinfo{pages}{24--33} (\bibinfo{year}{2019}).

\bibitem{ginzburg2016a}
\bibinfo{author}{{Ginzburg}, S.}, \bibinfo{author}{{Schlichting}, H.~E.} \&
  \bibinfo{author}{{Sari}, R.}
\newblock \bibinfo{title}{{Super-Earth Atmospheres: Self-consistent Gas
  Accretion and Retention}}.
\newblock \emph{\bibinfo{journal}{The Astrophysical Journal}}
  \textbf{\bibinfo{volume}{825}}, \bibinfo{pages}{29} (\bibinfo{year}{2016}).

\bibitem{Schlichting_Young_2022}
\bibinfo{author}{Schlichting, H.~E.} \& \bibinfo{author}{Young, E.~D.}
\newblock \bibinfo{title}{Chemical equilibrium between cores, mantles, and
  atmospheres of super-earths and sub-neptunes, and implications for their
  compositions, interiors and evolution}.
\newblock \emph{\bibinfo{journal}{Planetary Science Journal}}
  \textbf{\bibinfo{volume}{9}}, \bibinfo{pages}{19 pp.} (\bibinfo{year}{2022}).

\bibitem{Solomatov2009}
\bibinfo{author}{Solomatov, V.~S.}
\newblock \emph{\bibinfo{title}{Magma Oceans and Primordial Mantle
  Differentiation}}, vol.~\bibinfo{volume}{9} (\bibinfo{publisher}{Elsevier},
  \bibinfo{address}{Amsterdam}, \bibinfo{year}{2009}).

\bibitem{Young2019}
\bibinfo{author}{Young, E.} \emph{et~al.}
\newblock \bibinfo{title}{Near-equilibrium isotope fractionation during
  planetesimal evaporation}.
\newblock \emph{\bibinfo{journal}{Icarus}} \textbf{\bibinfo{volume}{323}},
  \bibinfo{pages}{1--15} (\bibinfo{year}{2019}).

\bibitem{Mukhopadhyay2012}
\bibinfo{author}{Mukhopadhyay, S.}
\newblock \bibinfo{title}{Early differentiation and volatile accretion recorded
  in deep-mantle neon and xenon}.
\newblock \emph{\bibinfo{journal}{Nature}} \textbf{\bibinfo{volume}{486}},
  \bibinfo{pages}{101--104} (\bibinfo{year}{2012}).

\bibitem{Mukhopadhyay2019}
\bibinfo{author}{Mukhopadhyay, S.} \& \bibinfo{author}{Parai, R.}
\newblock \bibinfo{title}{Noble gases: A record of earth's evolution and mantle
  dynamics}.
\newblock \emph{\bibinfo{journal}{Annual Review of Earth and Planetary
  Sciences}} \textbf{\bibinfo{volume}{47}}, \bibinfo{pages}{389--419}
  (\bibinfo{year}{2019}).

\bibitem{Williams2019}
\bibinfo{author}{Williams, C.~D.} \& \bibinfo{author}{Mukhopadhyay, S.}
\newblock \bibinfo{title}{Capture of nebular gases during earth's accretion is
  preserved in deep-mantle neon}.
\newblock \emph{\bibinfo{journal}{Nature}} \textbf{\bibinfo{volume}{565}},
  \bibinfo{pages}{78--81} (\bibinfo{year}{2019}).

\bibitem{Lammer2020}
\bibinfo{author}{Lammer, H.} \emph{et~al.}
\newblock \bibinfo{title}{Constraining the early evolution of {Venus} and
  {Earth} through atmospheric {Ar}, {Ne} isotope and bulk {K/U} ratios}.
\newblock \emph{\bibinfo{journal}{Icarus}} \textbf{\bibinfo{volume}{339}},
  \bibinfo{pages}{113551} (\bibinfo{year}{2020}).

\bibitem{Sharp2022}
\bibinfo{author}{Sharp, Z.} \& \bibinfo{author}{Olson, P.}
\newblock \bibinfo{title}{Multi-element constraints on the sources of volatiles
  to earth}.
\newblock \emph{\bibinfo{journal}{Geochimica et Cosmochimica Acta}}
  \textbf{\bibinfo{volume}{333}}, \bibinfo{pages}{124--135}
  (\bibinfo{year}{2022}).

\bibitem{Kurokawa2021}
\bibinfo{author}{Kurokawa, H.} \emph{et~al.}
\newblock \bibinfo{title}{Mars‚Äô atmospheric neon suggests
  volatile-rich primitive mantle}.
\newblock \emph{\bibinfo{journal}{Icarus}} \textbf{\bibinfo{volume}{370}},
  \bibinfo{pages}{114685} (\bibinfo{year}{2021}).

\bibitem{Peron2022}
\bibinfo{author}{Peron, S.} \& \bibinfo{author}{Mukhopadhyay, S.}
\newblock \bibinfo{title}{Krypton in the chassigny meteorite shows mars
  accreted chondritic volatiles before nebular gases}.
\newblock \emph{\bibinfo{journal}{Science}} \textbf{\bibinfo{volume}{377}},
  \bibinfo{pages}{320--324} (\bibinfo{year}{2022}).

\bibitem{Dauphas2011}
\bibinfo{author}{Dauphas, N.} \& \bibinfo{author}{Pourmand, A.}
\newblock \bibinfo{title}{Hf--w--th evidence for rapid growth of mars and its
  status as a planetary embryo}.
\newblock \emph{\bibinfo{journal}{Nature}} \textbf{\bibinfo{volume}{473}},
  \bibinfo{pages}{489--492} (\bibinfo{year}{2011}).

\bibitem{schlichting2014a}
\bibinfo{author}{{Schlichting}, H.~E.}
\newblock \bibinfo{title}{{Formation of Close in Super-Earths and
  Mini-Neptunes: Required Disk Masses and their Implications}}.
\newblock \emph{\bibinfo{journal}{The Astrophysical Journall}}
  \textbf{\bibinfo{volume}{795}}, \bibinfo{pages}{L15} (\bibinfo{year}{2014}).

\bibitem{Johansen2021}
\bibinfo{author}{Johansen, A.} \emph{et~al.}
\newblock \bibinfo{title}{A pebble accretion model for the formation of the
  terrestrial planets in the solar system}.
\newblock \emph{\bibinfo{journal}{Science Advances}}
  \textbf{\bibinfo{volume}{7}}, \bibinfo{pages}{eabc0444}
  (\bibinfo{year}{2021}).

\bibitem{Badro2015}
\bibinfo{author}{{Badro}, J.}, \bibinfo{author}{{Brodholt}, J.~P.},
  \bibinfo{author}{{Piet}, H.}, \bibinfo{author}{{Siebert}, J.} \&
  \bibinfo{author}{{Ryerson}, F.}
\newblock \bibinfo{title}{Core formation and core composition from coupled
  geochemical and geophysical constraints}.
\newblock \emph{\bibinfo{journal}{Proceedings of the National Academy of
  Sciences}} \textbf{\bibinfo{volume}{112}}, \bibinfo{pages}{12310--12314}
  (\bibinfo{year}{2015}).

\bibitem{Li2020_1038}
\bibinfo{author}{Li, Y.}, \bibinfo{author}{Vo{\v c}adlo, L.},
  \bibinfo{author}{Sun, T.} \& \bibinfo{author}{Brodholt, J.~P.}
\newblock \bibinfo{title}{The earth's core as a reservoir of water}.
\newblock \emph{\bibinfo{journal}{Nature Geoscience}}
  \textbf{\bibinfo{volume}{13}}, \bibinfo{pages}{453--458}
  (\bibinfo{year}{2020}).

\bibitem{Wood2008}
\bibinfo{author}{Wood, B.}
\newblock \bibinfo{title}{Accretion and core formation: Constraints from
  metal-silicate partitioning}.
\newblock \emph{\bibinfo{journal}{Philosophical Transactions - Royal Society of
  London, A}} \textbf{\bibinfo{volume}{366}}, \bibinfo{pages}{4339--4355}
  (\bibinfo{year}{2008}).

\bibitem{seager2007a}
\bibinfo{author}{{Seager}, S.}, \bibinfo{author}{{Kuchner}, M.},
  \bibinfo{author}{{Hier-Majumder}, C.~A.} \& \bibinfo{author}{{Militzer}, B.}
\newblock \bibinfo{title}{{Mass-Radius Relationships for Solid Exoplanets}}.
\newblock \emph{\bibinfo{journal}{The Astrophysical Journal}}
  \textbf{\bibinfo{volume}{669}}, \bibinfo{pages}{1279--1297}
  (\bibinfo{year}{2007}).

\bibitem{Anderson2001}
\bibinfo{author}{{Anderson}, O.~L.}, \bibinfo{author}{{Dubrovinsky}, L.},
  \bibinfo{author}{{Saxena}, S.~K.} \& \bibinfo{author}{{LeBihan}, T.}
\newblock \bibinfo{title}{{Experimental vibrational Gr{\"u}neisen ratio values
  for {\ensuremath{\in}}-iron up to 330 GPa at 300 K}}.
\newblock \emph{\bibinfo{journal}{Geophysical Research Letters}}
  \textbf{\bibinfo{volume}{28}}, \bibinfo{pages}{399--402}
  (\bibinfo{year}{2001}).

\bibitem{Kuwayama2020}
\bibinfo{author}{Kuwayama, Y.} \emph{et~al.}
\newblock \bibinfo{title}{Equation of state of liquid iron under extreme
  conditions}.
\newblock \emph{\bibinfo{journal}{Phys. Rev. Lett.}}
  \textbf{\bibinfo{volume}{124}}, \bibinfo{pages}{165701}
  (\bibinfo{year}{2020}).

\bibitem{Ikoma2006}
\bibinfo{author}{Ikoma, M.} \& \bibinfo{author}{Genda, H.}
\newblock \bibinfo{title}{Constraints on the mass of a habitable planet with
  water of nebular origin}.
\newblock \emph{\bibinfo{journal}{The Astrophysical Journal}}
  \textbf{\bibinfo{volume}{648}}, \bibinfo{pages}{696--706}
  (\bibinfo{year}{2006}).

\bibitem{Kite2021}
\bibinfo{author}{Kite, E.} \& \bibinfo{author}{Schaefer, L.}
\newblock \bibinfo{title}{Water on hot rocky exoplanets}.
\newblock \emph{\bibinfo{journal}{The Astrophysical Journal Letters}}
  \textbf{\bibinfo{volume}{909:L22}}, \bibinfo{pages}{6 pp.}
  (\bibinfo{year}{2021}).

\bibitem{Birch1964}
\bibinfo{author}{Birch, F.}
\newblock \bibinfo{title}{Density and composition of mantle and core}.
\newblock \emph{\bibinfo{journal}{Journal of Geophysical Research (1896-1977)}}
  \textbf{\bibinfo{volume}{69}}, \bibinfo{pages}{4377--4388}
  (\bibinfo{year}{1964}).

\bibitem{Umemoto2020}
\bibinfo{author}{Umemoto, K.} \& \bibinfo{author}{Hirose, K.}
\newblock \bibinfo{title}{Chemical compositions of the outer core examined by
  first principles calculations}.
\newblock \emph{\bibinfo{journal}{Earth and Planetary Science Letters}}
  \textbf{\bibinfo{volume}{531}}, \bibinfo{pages}{116009}
  (\bibinfo{year}{2020}).

\bibitem{Li2019}
\bibinfo{author}{Li, J.}, \bibinfo{author}{Chen, B.},
  \bibinfo{author}{Mookherjee, M.} \& \bibinfo{author}{Morard, G.}
\newblock \emph{\bibinfo{title}{Carbon versus Other Light Elements in
  Earth’s Core}}, \bibinfo{pages}{40–65}
  (\bibinfo{publisher}{Cambridge University Press}, \bibinfo{year}{2019}).

\bibitem{doyle2019a}
\bibinfo{author}{{Doyle}, A.~E.}, \bibinfo{author}{{Young}, E.~D.},
  \bibinfo{author}{{Klein}, B.}, \bibinfo{author}{{Zuckerman}, B.} \&
  \bibinfo{author}{{Schlichting}, H.~E.}
\newblock \bibinfo{title}{{Oxygen fugacities of extrasolar rocks: Evidence for
  an Earth-like geochemistry of exoplanets}}.
\newblock \emph{\bibinfo{journal}{Science}} \textbf{\bibinfo{volume}{366}},
  \bibinfo{pages}{356--359} (\bibinfo{year}{2019}).

\bibitem{Javoy2010}
\bibinfo{author}{Javoy, M.} \emph{et~al.}
\newblock \bibinfo{title}{The chemical composition of the earth: Enstatite
  chondrite models}.
\newblock \emph{\bibinfo{journal}{Earth and Planetary Science Letters}}
  \textbf{\bibinfo{volume}{293}}, \bibinfo{pages}{259--268}
  (\bibinfo{year}{2010}).

\bibitem{Dziewonski1981}
\bibinfo{author}{Dziewonski, A.~M.} \& \bibinfo{author}{Anderson, D.~L.}
\newblock \bibinfo{title}{Preliminary reference earth model}.
\newblock \emph{\bibinfo{journal}{Physics of the Earth and Planetary
  Interiors}} \textbf{\bibinfo{volume}{25}}, \bibinfo{pages}{297--356}
  (\bibinfo{year}{1981}).

\bibitem{Sanloup2004}
\bibinfo{author}{Sanloup, C.}, \bibinfo{author}{Fiquet, G.},
  \bibinfo{author}{Gregoryanz, E.}, \bibinfo{author}{Morard, G.} \&
  \bibinfo{author}{Mezouar, M.}
\newblock \bibinfo{title}{Effect of si on liquid fe compressibility:
  Implications for sound velocity in core materials}.
\newblock \emph{\bibinfo{journal}{Geophysical Research Letters}}
  \textbf{\bibinfo{volume}{31}} (\bibinfo{year}{2004}).

\bibitem{Umemoto2015}
\bibinfo{author}{Umemoto, K.} \& \bibinfo{author}{Hirose, K.}
\newblock \bibinfo{title}{Liquid iron-hydrogen alloys at outer core conditions
  by first-principles calculations}.
\newblock \emph{\bibinfo{journal}{Geophysical Research Letters}}
  \textbf{\bibinfo{volume}{42}}, \bibinfo{pages}{7513--7520}
  (\bibinfo{year}{2015}).

\bibitem{Kennett1995}
\bibinfo{author}{Kennett, B. L.~N.}, \bibinfo{author}{Engdahl, E.~R.} \&
  \bibinfo{author}{Buland, R.}
\newblock \bibinfo{title}{{Constraints on seismic velocities in the Earth from
  traveltimes}}.
\newblock \emph{\bibinfo{journal}{Geophysical Journal International}}
  \textbf{\bibinfo{volume}{122}}, \bibinfo{pages}{108--124}
  (\bibinfo{year}{1995}).

\bibitem{biersteker2021a}
\bibinfo{author}{{Biersteker}, J.~B.} \& \bibinfo{author}{{Schlichting}, H.~E.}
\newblock \bibinfo{title}{{Losing oceans: The effects of composition on the
  thermal component of impact-driven atmospheric loss}}.
\newblock \emph{\bibinfo{journal}{Monthly Notices of the Royal Astronomical
  Society}} \textbf{\bibinfo{volume}{501}}, \bibinfo{pages}{587--595}
  (\bibinfo{year}{2021}).

\bibitem{biersteker2019a}
\bibinfo{author}{{Biersteker}, J.~B.} \& \bibinfo{author}{{Schlichting}, H.~E.}
\newblock \bibinfo{title}{{Atmospheric mass-loss due to giant impacts: the
  importance of the thermal component for hydrogen-helium envelopes}}.
\newblock \emph{\bibinfo{journal}{Monthly Notices of the Royal Astronomical
  Society}} \textbf{\bibinfo{volume}{485}}, \bibinfo{pages}{4454--4463}
  (\bibinfo{year}{2019}).

\bibitem{Stahler2021}
\bibinfo{author}{St\"{a}hler, S.~C.} \emph{et~al.}
\newblock \bibinfo{title}{Seismic detection of the martian core}.
\newblock \emph{\bibinfo{journal}{Science}} \textbf{\bibinfo{volume}{373}},
  \bibinfo{pages}{443--448} (\bibinfo{year}{2021}).

\bibitem{Brasser2013}
\bibinfo{author}{Brasser, R.}
\newblock \bibinfo{title}{The formation of mars: Building blocks and accretion
  time scale}.
\newblock \emph{\bibinfo{journal}{Space Science Reviews}}
  \textbf{\bibinfo{volume}{174}}, \bibinfo{pages}{11--25}
  (\bibinfo{year}{2013}).

\bibitem{Benz2007}
\bibinfo{author}{Benz, W.}, \bibinfo{author}{Anic, A.},
  \bibinfo{author}{Horner, J.} \& \bibinfo{author}{Whitby, J.~A.}
\newblock \bibinfo{title}{The origin of mercury}.
\newblock \emph{\bibinfo{journal}{Space Science Reviews}}
  \textbf{\bibinfo{volume}{132}}, \bibinfo{pages}{189--202}
  (\bibinfo{year}{2007}).

\bibitem{Riner2008}
\bibinfo{author}{Riner, M.~A.}, \bibinfo{author}{Bina, C.~R.},
  \bibinfo{author}{Robinson, M.~S.} \& \bibinfo{author}{Desch, S.~J.}
\newblock \bibinfo{title}{Internal structure of mercury: Implications of a
  molten core}.
\newblock \emph{\bibinfo{journal}{Journal of Geophysical Research: Planets}}
  \textbf{\bibinfo{volume}{113}}.

\bibitem{Margot2018}
\bibinfo{author}{Margot, J.-L.}, \bibinfo{author}{Hauck, S.~A.},
  \bibinfo{author}{Mazarico, E.}, \bibinfo{author}{Padovan, S.} \&
  \bibinfo{author}{Peale, S.~J.}
\newblock \emph{\bibinfo{title}{Mercury’s Internal Structure}},
  \bibinfo{pages}{85–113}.
\newblock Cambridge Planetary Science (\bibinfo{publisher}{Cambridge University
  Press}, \bibinfo{year}{2018}).

\bibitem{Corgne2008}
\bibinfo{author}{Corgne, A.}, \bibinfo{author}{Keshav, S.},
  \bibinfo{author}{Wood, B.~J.}, \bibinfo{author}{McDonough, W.~F.} \&
  \bibinfo{author}{Fei, Y.}
\newblock \bibinfo{title}{Metal-silicate partitioning and constraints on core
  composition and oxygen fugacity during earth accretion}.
\newblock \emph{\bibinfo{journal}{Geochimica et Cosmochimica Acta}}
  \textbf{\bibinfo{volume}{72}}, \bibinfo{pages}{574--589}
  (\bibinfo{year}{2008}).

\bibitem{Fegley1987}
\bibinfo{author}{{Fegley}, J., Bruce} \& \bibinfo{author}{{Cameran}, A. G.~W.}
\newblock \bibinfo{title}{A vaporization model for iron/silicate fractionation
  in the mercury protoplanet}.
\newblock \emph{\bibinfo{journal}{Earth and Planetary Science Letters}}
  \textbf{\bibinfo{volume}{82}}, \bibinfo{pages}{207--222}
  (\bibinfo{year}{1987}).

\bibitem{schaefer2009a}
\bibinfo{author}{{Schaefer}, L.} \& \bibinfo{author}{{Fegley}, B.}
\newblock \bibinfo{title}{{Chemistry of Silicate Atmospheres of Evaporating
  Super-Earths}}.
\newblock \emph{\bibinfo{journal}{The Astrophysical Journall}}
  \textbf{\bibinfo{volume}{703}}, \bibinfo{pages}{L113--L117}
  (\bibinfo{year}{2009}).

\bibitem{Hastie1986}
\bibinfo{author}{Hastie, J.} \& \bibinfo{author}{Bonnell, D.}
\newblock \bibinfo{title}{{A predictive thermodynamic model of oxide and halide
  glass phase equilibria}}.
\newblock \emph{\bibinfo{journal}{Journal of Non-Crystalline Solids}}
  \textbf{\bibinfo{volume}{84}}, \bibinfo{pages}{151--158}
  (\bibinfo{year}{1986}).

\bibitem{Xiang_1997}
\bibinfo{author}{{Xiang}, Y.}, \bibinfo{author}{{Sun}, D.},
  \bibinfo{author}{{Fan}, W.} \& \bibinfo{author}{{Gong}, X.}
\newblock \bibinfo{title}{Generalized simulated annealing algorithm and its
  application to the thomson model}.
\newblock \emph{\bibinfo{journal}{Physics Letters A}}
  \textbf{\bibinfo{volume}{233}}, \bibinfo{pages}{216--220}
  (\bibinfo{year}{1997}).

\bibitem{Foreman2019}
\bibinfo{author}{Foreman-Mackey, D.} \emph{et~al.}
\newblock \bibinfo{title}{{emcee v3: A Python ensemble sampling toolkit for
  affine-invariant MCMC}}.
\newblock \emph{\bibinfo{journal}{The Journal of Open Source Software}}
  \textbf{\bibinfo{volume}{43}} (\bibinfo{year}{2019}).

\bibitem{Goodman2010}
\bibinfo{author}{Goodman, J.} \& \bibinfo{author}{Weare, J.}
\newblock \bibinfo{title}{{Ensemble samplers with affine invariance}}.
\newblock \emph{\bibinfo{journal}{Communications in Applied Mathematics and
  Computational Science}} \textbf{\bibinfo{volume}{5}}, \bibinfo{pages}{65 --
  80} (\bibinfo{year}{2010}).

\bibitem{Kovacevic_2022}
\bibinfo{author}{Kova{\v c}evi{\'c}, T.},
  \bibinfo{author}{Gonz{\'a}lez-Cataldo, F.}, \bibinfo{author}{Stewart, S.~T.}
  \& \bibinfo{author}{Militzer, B.}
\newblock \bibinfo{title}{Miscibility of rock and ice in the interiors of water
  worlds}.
\newblock \emph{\bibinfo{journal}{Scientific Reports}}
  \textbf{\bibinfo{volume}{12}}, \bibinfo{pages}{13055} (\bibinfo{year}{2022}).

\bibitem{Xiao2018}
\bibinfo{author}{Xiao, B.} \& \bibinfo{author}{Stixrude, L.}
\newblock \bibinfo{title}{Critical vaporization of {MgSiO$_3$}}.
\newblock \emph{\bibinfo{journal}{Proceedings of the National Academy of
  Sciences}} \textbf{\bibinfo{volume}{115}}, \bibinfo{pages}{5371--5376}
  (\bibinfo{year}{2018}).

\bibitem{Boorstein1974}
\bibinfo{author}{Boorstein, W.~M.} \& \bibinfo{author}{Pehlke, R.~D.}
\newblock \bibinfo{title}{Measurement of hydrogen solubility in liquid iron
  alloys employing a constant volume technique}.
\newblock \emph{\bibinfo{journal}{Metallurgical and Materials Transactions B}}
  \textbf{\bibinfo{volume}{5}}, \bibinfo{pages}{399--405}
  (\bibinfo{year}{1974}).

\bibitem{Waseda_2012}
\bibinfo{author}{Waseda, Y.}
\newblock \bibinfo{title}{Interaction parameters in metallic solutions
  estimated from liquid structure and the heat of solution at infinite
  dilution}.
\newblock \emph{\bibinfo{journal}{High Temperature Materials and Processes}}
  \textbf{\bibinfo{volume}{31}}, \bibinfo{pages}{203--208}
  (\bibinfo{year}{2012}).

\bibitem{Righter_2020}
\bibinfo{author}{{Righter, K and Rowland, R. II and Yang, S. and Humayun, M.}}
\newblock \bibinfo{title}{Activity coefficients of siderophile elements in
  fe-si liquids at high pressure}.
\newblock \emph{\bibinfo{journal}{Geochemical Perspectives Letters}}
  \textbf{\bibinfo{volume}{15}}, \bibinfo{pages}{44--49}
  (\bibinfo{year}{2020}).

\bibitem{Hirschmann2012}
\bibinfo{author}{Hirschmann, M.}, \bibinfo{author}{Withers, A.},
  \bibinfo{author}{Ardia, P.} \& \bibinfo{author}{Foley, N.}
\newblock \bibinfo{title}{Solubility of molecular hydrogen in silicate melts
  and consequences for volatile evolution of terrestrial planets}.
\newblock \emph{\bibinfo{journal}{Earth and Planetary Science Letters}}
  \textbf{\bibinfo{volume}{345}}, \bibinfo{pages}{38--48}
  (\bibinfo{year}{2012}).

\bibitem{Okuchi1997}
\bibinfo{author}{{Okuchi}, T.}
\newblock \bibinfo{title}{{Hydrogen partitioning into molten iron at high
  pressure: implications for Earth's core}}.
\newblock \emph{\bibinfo{journal}{Science}} \textbf{\bibinfo{volume}{278}},
  \bibinfo{pages}{1781--1784} (\bibinfo{year}{1997}).

\bibitem{Moore1998}
\bibinfo{author}{Moore, G.}, \bibinfo{author}{Vennemann, T.} \&
  \bibinfo{author}{Carmichael, I.}
\newblock \bibinfo{title}{{An empirical model for the solubility of H$_2$O in
  magmas to 3 kilobars}}.
\newblock \emph{\bibinfo{journal}{American Mineralogist}}
  \textbf{\bibinfo{volume}{83}}, \bibinfo{pages}{36--42}
  (\bibinfo{year}{1998}).

\bibitem{Karki2000}
\bibinfo{author}{{Karki}, B.~B.}, \bibinfo{author}{{Wentzcovitch}, R.~M.},
  \bibinfo{author}{{de Gironcoli}, S.} \& \bibinfo{author}{{Baroni}, S.}
\newblock \bibinfo{title}{{Ab initio lattice dynamics of {MgSiO$_{3}$}
  perovskite at high pressure}}.
\newblock \emph{\bibinfo{journal}{Physical Review B: Condensed Matter and
  Materials Physics}} \textbf{\bibinfo{volume}{62}},
  \bibinfo{pages}{14750--14756} (\bibinfo{year}{2000}).

\bibitem{Katsura2009}
\bibinfo{author}{Katsura, T.} \emph{et~al.}
\newblock \bibinfo{title}{Thermal expansion of forsterite at high pressures
  determined by in situ x-ray diffraction: The adiabatic geotherm in the upper
  mantle}.
\newblock \emph{\bibinfo{journal}{Physics of the Earth and Planetary
  Interiors}} \textbf{\bibinfo{volume}{174}}, \bibinfo{pages}{86--92}
  (\bibinfo{year}{2009}).

\bibitem{Shahar2020}
\bibinfo{author}{Shahar, A.} \& \bibinfo{author}{Young, E.~D.}
\newblock \bibinfo{title}{An assessment of iron isotope fractionation during
  core formation}.
\newblock \emph{\bibinfo{journal}{Chemical Geology}}
  \textbf{\bibinfo{volume}{554}}, \bibinfo{pages}{119800}
  (\bibinfo{year}{2020}).

\bibitem{Young2015}
\bibinfo{author}{Young, E.~D.} \emph{et~al.}
\newblock \bibinfo{title}{High-temperature equilibrium isotope fractionation of
  non-traditional stable isotopes: Experiments, theory, and applications}.
\newblock \emph{\bibinfo{journal}{Chemical Geology}}
  \textbf{\bibinfo{volume}{395}}, \bibinfo{pages}{176--195}
  (\bibinfo{year}{2015}).

\bibitem{Hallis2015}
\bibinfo{author}{Hallis, L.~J.} \emph{et~al.}
\newblock \bibinfo{title}{Evidence for primordial water in earth\&\#x2019;s
  deep mantle}.
\newblock \emph{\bibinfo{journal}{Science}} \textbf{\bibinfo{volume}{350}},
  \bibinfo{pages}{795--797} (\bibinfo{year}{2015}).

\bibitem{Sossi2016}
\bibinfo{author}{Sossi, P.~A.}, \bibinfo{author}{Nebel, O.} \&
  \bibinfo{author}{Foden, J.}
\newblock \bibinfo{title}{Iron isotope systematics in planetary reservoirs}.
\newblock \emph{\bibinfo{journal}{Earth and Planetary Science Letters}}
  \textbf{\bibinfo{volume}{452}}, \bibinfo{pages}{295--308}
  (\bibinfo{year}{2016}).

\bibitem{Pringle2014}
\bibinfo{author}{Pringle, E.~A.}, \bibinfo{author}{Moynier, F.},
  \bibinfo{author}{Savage, P.~S.}, \bibinfo{author}{Badro, J.} \&
  \bibinfo{author}{Barrat, J.-A.}
\newblock \bibinfo{title}{Silicon isotopes in angrites and volatile loss in
  planetesimals}.
\newblock \emph{\bibinfo{journal}{Proceedings of the National Academy of
  Sciences}} \textbf{\bibinfo{volume}{111}}, \bibinfo{pages}{17029--17032}
  (\bibinfo{year}{2014}).

\bibitem{Savage2012}
\bibinfo{author}{Savage, P.~S.} \& \bibinfo{author}{Moynier, F.}
\newblock \bibinfo{title}{Silicon isotopic variation in enstatite meteorites:
  Clues to their origin and earth-forming material}.
\newblock \emph{\bibinfo{journal}{Earth and Planetary Science Letters}}
  \textbf{\bibinfo{volume}{361}}, \bibinfo{pages}{487--496}
  (\bibinfo{year}{2013}).

\bibitem{Geiss2003}
\bibinfo{author}{Geiss, J.} \& \bibinfo{author}{Gloeckler, G.}
\newblock \bibinfo{title}{Isotopic composition of h, he and ne in the
  protosolar cloud}.
\newblock \emph{\bibinfo{journal}{Space Science Reviews}}
  \textbf{\bibinfo{volume}{106}}, \bibinfo{pages}{3--18}
  (\bibinfo{year}{2003}).

\bibitem{Altwegg2015}
\bibinfo{author}{Altwegg, K.} \emph{et~al.}
\newblock \bibinfo{title}{{67P/Churyumov-Gerasimenko}, a {Jupiter} family comet
  with a high {D/H} ratio}.
\newblock \emph{\bibinfo{journal}{Science}} \textbf{\bibinfo{volume}{347}},
  \bibinfo{pages}{1261952} (\bibinfo{year}{2015}).

\bibitem{BockeleeMorvan2012}
\bibinfo{author}{{Bockel\'ee-Morvan, D.}} \emph{et~al.}
\newblock \bibinfo{title}{Herschel measurements of the {D/H} and
  {$^{16}$O/$^{18}$O} ratios in water in the {Oort-cloud} comet {C/2009P1}
  ({Garradd})}.
\newblock \emph{\bibinfo{journal}{Astronomy \& Astrophysics}}
  \textbf{\bibinfo{volume}{544}}, \bibinfo{pages}{L15} (\bibinfo{year}{2012}).

\bibitem{Alexander2012}
\bibinfo{author}{Alexander, C. M.~O.} \emph{et~al.}
\newblock \bibinfo{title}{The provenances of asteroids, and their contributions
  to the volatile inventories of the terrestrial planets}.
\newblock \emph{\bibinfo{journal}{Science}} \textbf{\bibinfo{volume}{337}},
  \bibinfo{pages}{721--723} (\bibinfo{year}{2012}).

\bibitem{Waite2009}
\bibinfo{author}{Waite~Jr, J.~H.} \emph{et~al.}
\newblock \bibinfo{title}{Liquid water on {Enceladus} from observations of
  ammonia and {$^{40}$Ar} in the plume}.
\newblock \emph{\bibinfo{journal}{Nature}} \textbf{\bibinfo{volume}{460}},
  \bibinfo{pages}{487--490} (\bibinfo{year}{2009}).

\bibitem{Gough1980}
\bibinfo{author}{Gough, D.~O.}
\newblock \bibinfo{title}{Solar interior structure and luminosity variations}.
\newblock \emph{\bibinfo{journal}{Solar Physics}}
  \textbf{\bibinfo{volume}{74}}, \bibinfo{pages}{21--34}
  (\bibinfo{year}{1981}).

\bibitem{Genda2008}
\bibinfo{author}{Genda, H.} \& \bibinfo{author}{Ikoma, M.}
\newblock \bibinfo{title}{Origin of the ocean on the earth: Early evolution of
  water d/h in a hydrogen-rich atmosphere}.
\newblock \emph{\bibinfo{journal}{Icarus}} \textbf{\bibinfo{volume}{194}},
  \bibinfo{pages}{42--52} (\bibinfo{year}{2008}).

\bibitem{Lammer2014}
\bibinfo{title}{{Origin and loss of nebula-captured hydrogen envelopes from
  `sub'- to `super-Earths' in the habitable zone of Sun-like stars}}.
\newblock \emph{\bibinfo{journal}{Monthly Notices of the Royal Astronomical
  Society}} \textbf{\bibinfo{volume}{439}}, \bibinfo{pages}{3225--3238}
  (\bibinfo{year}{2014}).

\bibitem{Yang2013}
\bibinfo{author}{Yang, L.}, \bibinfo{author}{Ciesla, F.~J.} \&
  \bibinfo{author}{Alexander, C.~M.}
\newblock \bibinfo{title}{The d/h ratio of water in the solar nebula during its
  formation and evolution}.
\newblock \emph{\bibinfo{journal}{Icarus}} \textbf{\bibinfo{volume}{226}},
  \bibinfo{pages}{256--267} (\bibinfo{year}{2013}).

\bibitem{Holton1990}
\bibinfo{author}{Holton, J.~R.}
\newblock \bibinfo{title}{On the global exchange of mass between the
  stratosphere and troposphere}.
\newblock \emph{\bibinfo{journal}{Journal of Atmospheric Sciences}}
  \textbf{\bibinfo{volume}{47}}, \bibinfo{pages}{392--395}
  (\bibinfo{year}{1990}).

\bibitem{Pahlevan2019}
\bibinfo{author}{Pahlevan, K.}, \bibinfo{author}{Schaefer, L.} \&
  \bibinfo{author}{Hirschmann, M.~M.}
\newblock \bibinfo{title}{Hydrogen isotopic evidence for early oxidation of
  silicate earth}.
\newblock \emph{\bibinfo{journal}{Earth and Planetary Science Letters}}
  \textbf{\bibinfo{volume}{526}}, \bibinfo{pages}{115770}
  (\bibinfo{year}{2019}).

\bibitem{Krasnopolsky1998}
\bibinfo{author}{Krasnopolsky, V.~A.}, \bibinfo{author}{Mumma, M.~J.} \&
  \bibinfo{author}{Gladstone, G.~R.}
\newblock \bibinfo{title}{Detection of atomic deuterium in the upper atmosphere
  of mars}.
\newblock \emph{\bibinfo{journal}{Science}} \textbf{\bibinfo{volume}{280}},
  \bibinfo{pages}{1576--1580} (\bibinfo{year}{1998}).

\bibitem{Young2014}
\bibinfo{author}{Young, E.}, \bibinfo{author}{Yeung, L.} \&
  \bibinfo{author}{Kohl, I.}
\newblock \bibinfo{title}{On the Œî17o budget of atmospheric o2}.
\newblock \emph{\bibinfo{journal}{Geochimica et Cosmochimica Acta}}
  \textbf{\bibinfo{volume}{135}}, \bibinfo{pages}{102‚Äì125}
  (\bibinfo{year}{2014}).

\bibitem{Yeung2017}
\bibinfo{author}{Yeung, L.~Y.} \emph{et~al.}
\newblock \bibinfo{title}{Extreme enrichment in atmospheric 15n15n}.
\newblock \emph{\bibinfo{journal}{Science advances}}
  \textbf{\bibinfo{volume}{3}}, \bibinfo{pages}{eaao6741}
  (\bibinfo{year}{2017}).

\bibitem{Barnes2014}
\bibinfo{author}{Barnes, J.~J.} \emph{et~al.}
\newblock \bibinfo{title}{The origin of water in the primitive moon as revealed
  by the lunar highlands samples}.
\newblock \emph{\bibinfo{journal}{Earth and Planetary Science Letters}}
  \textbf{\bibinfo{volume}{390}}, \bibinfo{pages}{244--252}
  (\bibinfo{year}{2014}).

\bibitem{Lellouch2001}
\bibinfo{author}{Lellouch, E.} \emph{et~al.}
\newblock \bibinfo{title}{The deuterium abundance in jupiter and saturn from
  iso-sws observations}.
\newblock \emph{\bibinfo{journal}{Astronomy \& Astrophysics}}
  \textbf{\bibinfo{volume}{670}}, \bibinfo{pages}{610--622}
  (\bibinfo{year}{2001}).

\bibitem{Feuchtgruber2013}
\bibinfo{author}{Feuchtgruber, H.} \emph{et~al.}
\newblock \bibinfo{title}{The d/h ratio in the atmospheres of uranus and
  neptune from herschel-pacs observations}.
\newblock \emph{\bibinfo{journal}{Astronomy \& Astrophysics}}
  \textbf{\bibinfo{volume}{551}}, \bibinfo{pages}{A126} (\bibinfo{year}{2013}).

\end{thebibliography}

\end{singlespace}

\noindent {\bfseries Acknowledgments}  This AEThER publication is funded in part by the Alfred P. Sloan Foundation under grant G202114194.  H.E.S. gratefully acknowledges NASA grant 80NSSC18K0828 for financial support during preparation and submission of this work. 
\\

\noindent
{\bfseries Author Contribution} All authors contributed significantly to the formal analysis. E.D.Y. led the writing of the paper aided by A.S. and H.E.S..
\\

\noindent
{\bfseries Author Information} The authors declare that they have no
competing financial interests. Correspondence and requests for materials
should be addressed to E.D.Y.~(email: eyoung@epss.ucla.edu).\\

\clearpage
\begin{singlespace}
\noindent{\bf Extended Data}

\renewcommand{\figurename}{Extended Data Figure }
\renewcommand{\thefigure}{1}
 \begin{figure}[!h]
\centering
\includegraphics[width=5.8 in]{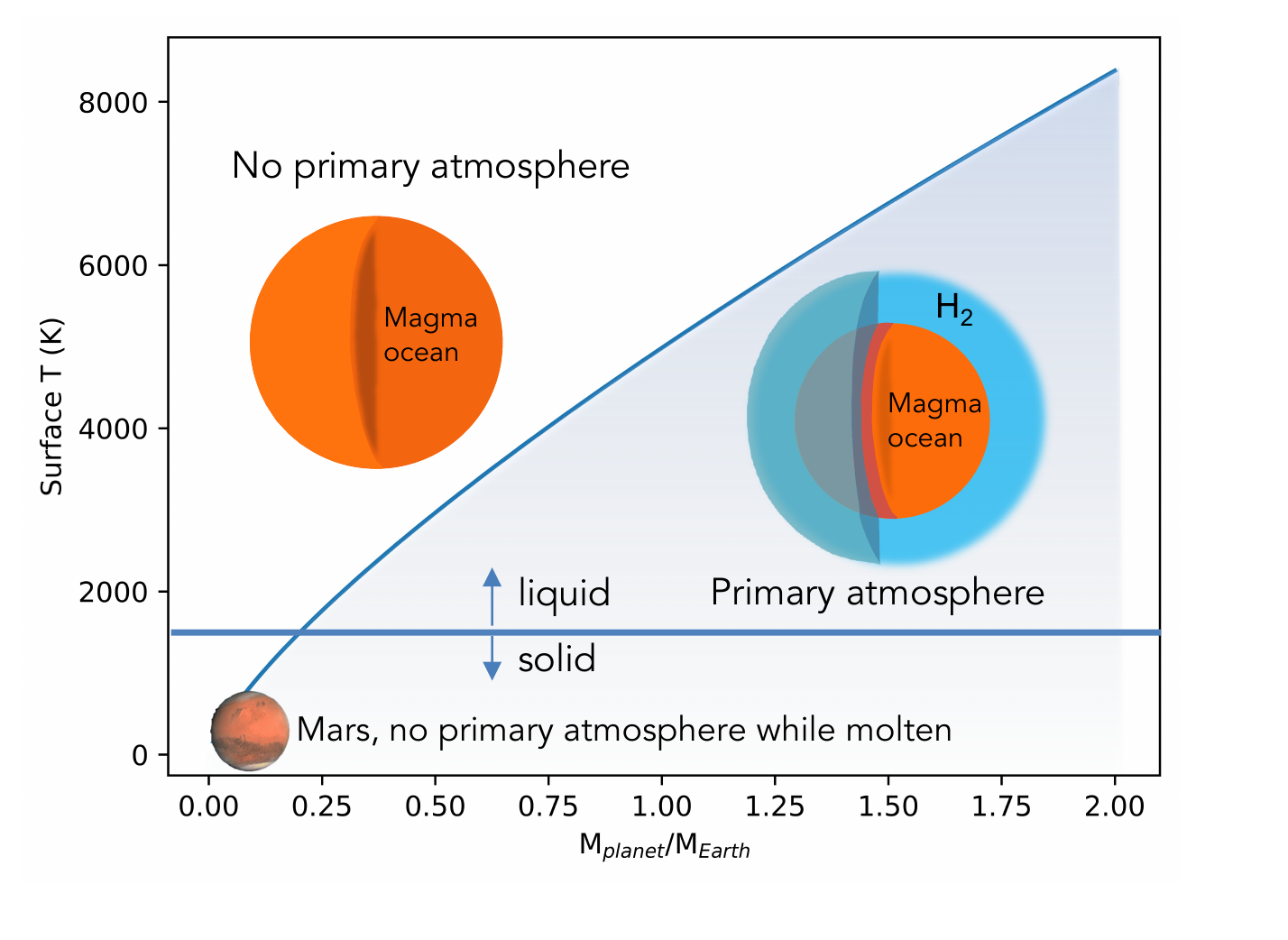}
\caption{Plot of surface temperatures that allow for a gravitationally bound primary H$_2$ atmosphere versus mass of planetary embryos \cite{ginzburg2016a}.  The region in temperature-mass space where a primary atmosphere is possible is shaded. The approximate solidus for silicate melt is overlain as the horizontal line. }
\end{figure}

\renewcommand{\figurename}{Extended Data Figure }
\renewcommand{\thefigure}{2}
 \begin{figure}[!h]
\centering
\includegraphics[width=5.8 in]{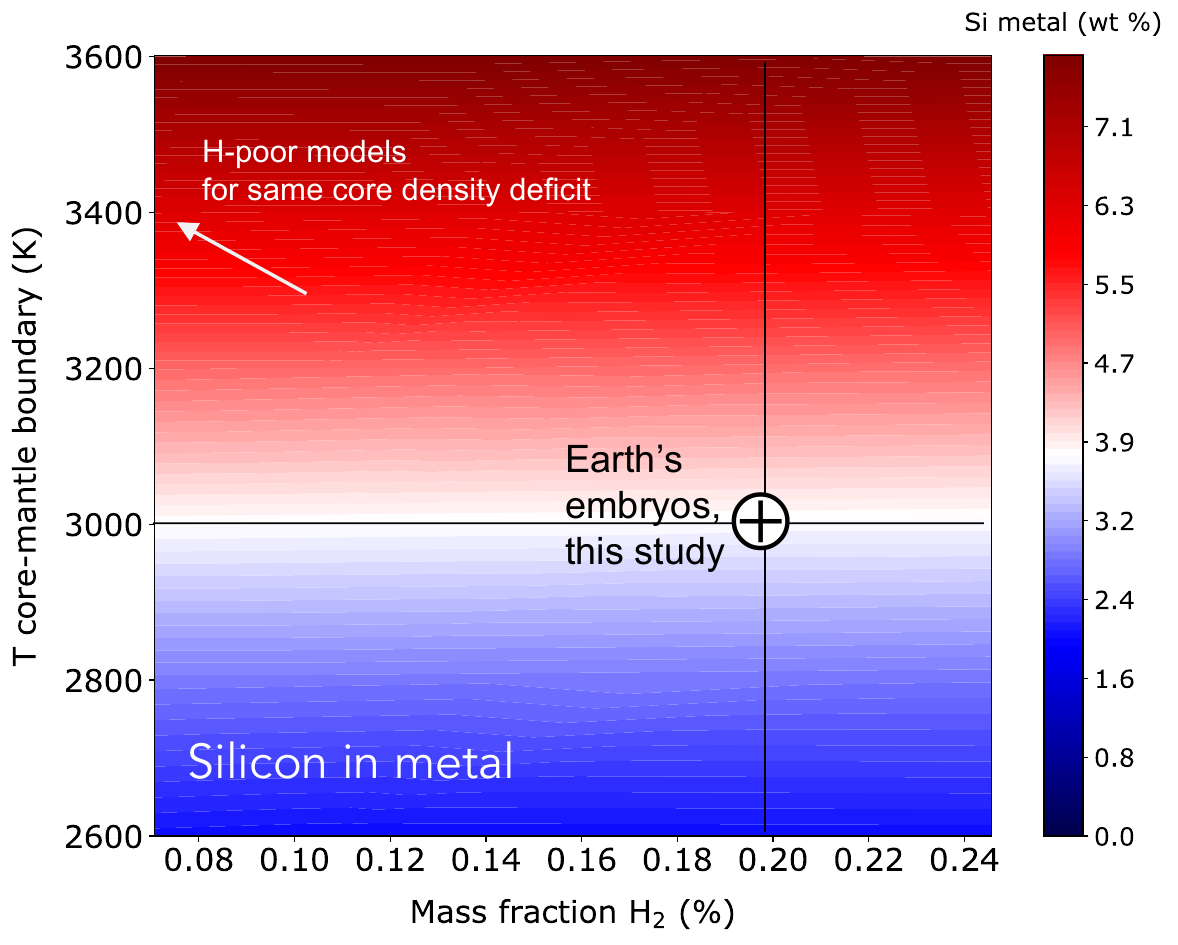}
\caption{Summary of equilibrium calculations for Si in metal in embryos with masses of $0.5 M_\oplus$ as a function of metal-silicate equilibration temperature (T$_{\rm core-mantle}$) and mass fraction of initial primary H$_2$-rich atmosphere relative to the planet. Arrow illustrates values for models that satisfy the required density deficit in the core but where H is scarce or absent. }
\end{figure}

\renewcommand{\figurename}{Extended Data Figure }
\renewcommand{\thefigure}{3}
 \begin{figure}[!h]
\centering
\includegraphics[width=5.8 in]{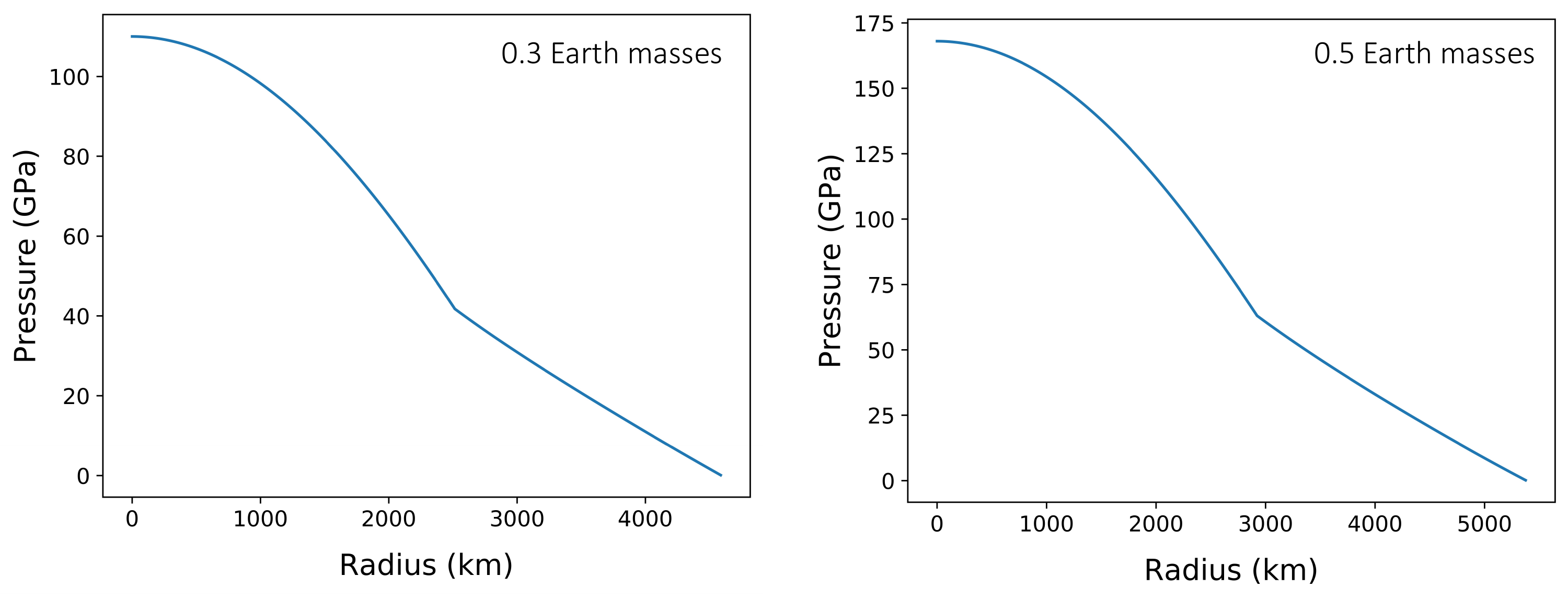}
\caption{Pressure vs.\ radius for $0.3 M_{\oplus}$ (left) and $0.5 M_{\oplus}$ (right) embryos. Breaks in slope mark the core-mantle boundaries.  }
\end{figure}

\renewcommand{\figurename}{Extended Data Figure }
\renewcommand{\thefigure}{4}
 \begin{figure}[!h]
\centering
\includegraphics[width=5.8 in]{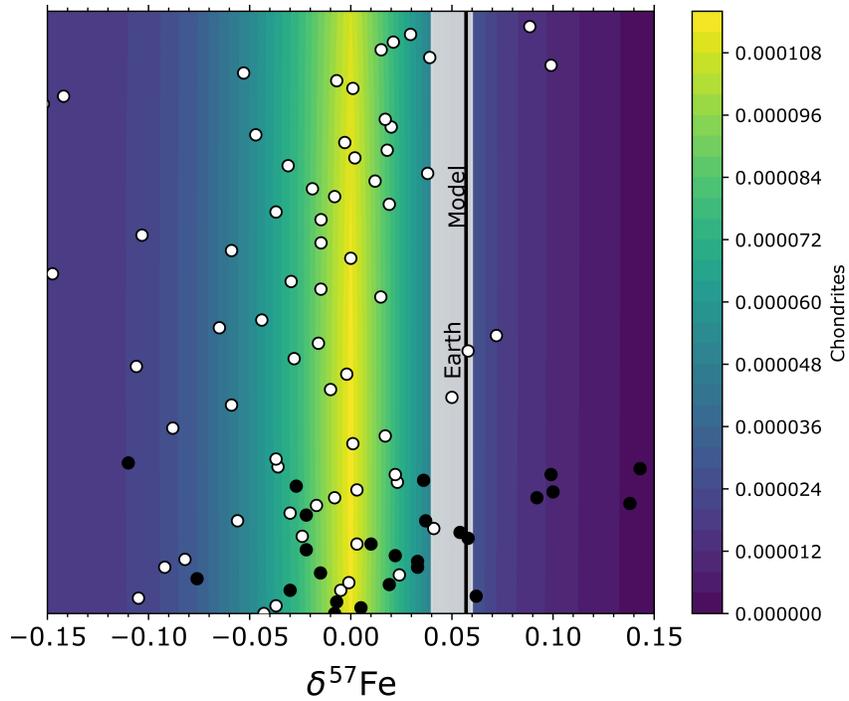}
\caption{ Iron isotope ratios of bulk silicate for model embryos compared with recent estimates for bulk Earth (grey bar) and chondrites showing that the model reproduces the offset between initial materials (chondrites, $\delta^{57}$Fe $=0$) and Earth. Black filled symbols are for E chondrites while white symbols are for all other chondrite groups.  The multi-color contours are probability densities for the chondrite $\delta^{57}$Fe values with an average indistinguishable from $\delta^{57}$Fe $= 0.0$.   The y axis values are assigned randomly to each datum in equal spacings for clarity, with E chondrites confined to the lower quarter of the ordinate.   }
\end{figure}

\renewcommand{\figurename}{Extended Data Figure }
\renewcommand{\thefigure}{5}
 \begin{figure}[!h]
\centering
\includegraphics[width=5.8 in]{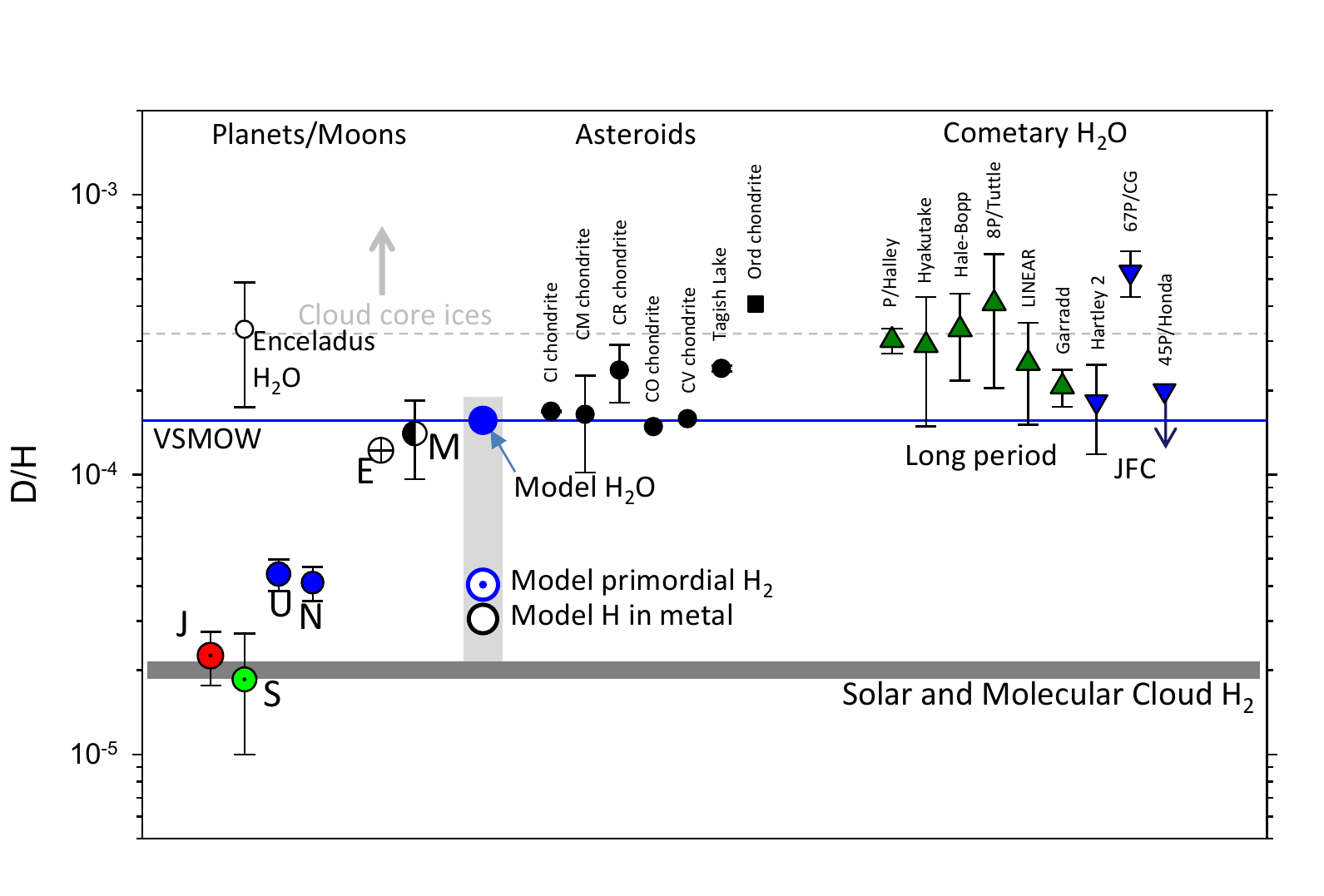}
\caption{ Summary of D/H ratios for Solar System materials from a variety of literature sources.  The circle + symbol labelled E denotes bulk Earth\cite{Hallis2015}. Black/white symbol labelled M refers to lunar highland apatites\cite{Barnes2014}.  Blue symbols refer to calculated values for original water based on measured asteroidal rock values \cite{Alexander2012}. U, N, J, and S refer to Uranus, Neptune, Jupiter, and Saturn, respectively\cite{Lellouch2001,Feuchtgruber2013}.  Model values for Earth's water, primordial hydrogen atmosphere, and metal described here are indicated within the grey box, where the model is assigned the terrestrial value. }
\end{figure}

\end{singlespace}

\end{document}